\documentclass[a4paper]{article}

\usepackage{authblk}
\usepackage[margin=1in]{geometry}
\usepackage{float}
\usepackage{graphicx}%
\usepackage{multirow}%
\usepackage{amsmath,amssymb,amsfonts}%
\usepackage{amsthm}%
\usepackage{mathrsfs}%
\usepackage[title]{appendix}%
\usepackage{xcolor}%
\usepackage{textcomp}%
\usepackage{ulem}
\usepackage{manyfoot}%
\usepackage{booktabs}%
\usepackage{algorithm}%
\usepackage{algorithmicx}%
\usepackage{algpseudocode}%
\usepackage{listings}%
\usepackage{subcaption}
\usepackage{enumitem}
\usepackage{adjustbox}
\usepackage{braket}
\usepackage{tikz}
\usetikzlibrary{quantikz2}
\title{\textbf{Quantum algorithms for second-order boundary value problems}}

\author[1]{Lauri Kettunen}
\author[1]{Tejas Shinde}

\affil[1]{University of Jyväskylä, Faculty of Information Technology, P.O.Box 35, FI-40014 Jyväskylä, Finland}

\begin{document}

% \articletype{Article} %	 e.g. Paper, Letter, Topical Review...
\maketitle

\begin{abstract}
\normalsize
Second-order boundary value problems are central to computational science, yet standard matrix-based numerical formulations can obscure the local geometric structure that quantum circuits may exploit. Here we introduce a framework for explicitly constructing finite-dimensional counterparts of continuous differential operators in a form compatible with quantum computation. Starting from the exterior derivative, its adjoint, and the Hodge operator, we derive discrete realizations of second-order operators on primal--dual cell complexes and reformulate them as star-local update rules expressed through explicit functions that return the relevant bounding chains rather than through matrix representations. This yields simple, uniform, and scalable quantum circuits. We demonstrate the construction for div--grad and curl--curl operators, showing that the same star-local compilation principle extends across different operators, cell complexes, and manifold dimensions within a common framework. More generally, the framework provides a systematic route to quantum algorithms for partial differential equations.
\end{abstract}


\section{Introduction}

Second-order boundary value problems are among the most important computational models in science and engineering. Elliptic, Poisson-type, and curl--curl equations arise in applications ranging from electromagnetism and fluid dynamics to materials modelling and multiphysics simulation. Classical numerical approaches, including finite element methods \cite{Ciarlet1978} and finite difference methods \cite{Yee1966,BossavitPIER2001}, approximate the solutions of these continuous problems in finite-dimensional spaces. This leads to finite systems of equations whose solution often dominates the computational cost in large-scale simulations.

Quantum computing has stimulated considerable interest in alternative approaches to partial differential equations and boundary value problems. Quantum linear-system algorithms form one of the central algorithmic paradigms in quantum scientific computing and have led to extensive research on quantum methods for large-scale numerical problems \cite{Montanaro2016}. Many existing quantum algorithms for differential equations start from a matrix representation of a discretized problem. Examples include quantum linear-system algorithms \cite{HHL2009,Clader2013,Childs2017QLSA}, more recent quantum algorithms for partial differential equations based on finite-difference and spectral discretizations \cite{Childs2021PDE}, and related approaches surveyed in \cite{Costa2022Review}. In these matrix-based approaches, the discretized operator is typically implemented through block encoding, Hamiltonian simulation, or related quantum primitives. Other approaches, such as quantum-walk and quantum lattice Boltzmann methods, instead exploit local propagation or shift-based circuit primitives \cite{Budinski2021ADE,bastida2026quantum}.

Starting from an assembled matrix is natural from the viewpoint of numerical linear algebra. An alternative is to directly construct finite-dimensional counterparts of the differential structures from which the boundary value problem is built. For example, the familiar operators ${\rm grad}$, ${\rm curl}$, and ${\rm div}$ arise as metric realizations of the exterior derivative ${\mathrm d}$ on differential forms \cite{Bossavit-Japan2}. Second-order operators are then obtained from this first-order differential structure together with the Hodge operator and the adjoint exterior derivative \cite{Flanders,Baez-Muniain,Frankel}. In particular, up to sign convention, the {\em Laplace--de~Rham operator} on differential forms is
\[
{\mathrm d}\delta+\delta{\mathrm d},
\]
where $\delta$ is the adjoint of ${\mathrm d}$. The div--grad and curl--curl operators considered here are particular metric representatives of this general exterior-calculus construction: depending on the form degree and gauge or constraint conditions, they correspond to one of its constituent second-order parts. This observation motivates an alternative route for quantum algorithms: rather than taking a globally assembled matrix as the primary object, the algorithm can be constructed directly from the local differential structure from which the operator is built.

We implement this structure-first viewpoint as follows. Unlike quantum lattice-dynamics approaches, the construction does not first replace the PDE by an auxiliary local propagation model; instead, the circuit locality is derived directly from the exterior derivative, its adjoint, and the Hodge-type constitutive structure. In a finite-dimensional setting, differentiation requires a replacement for the virtual neighbourhoods appearing in the continuous definition of ${\mathrm d}$; this role is played by local incidence relations on cell complexes \cite{Tarhasaari-Kettunen-Bossavit1999,Bossavit-Kettunen1999}. Starting from the exterior derivative, its adjoint, and Hodge-type constitutive operators, we construct finite-dimensional counterparts of second-order differential operators on primal--dual cell complexes. The support of the resulting operator is determined by the {\em star of a cell} \cite{Munkres1984}, which contains precisely the neighbouring entities needed for evaluating local differential interactions. This locality is therefore not imposed as an external propagation rule, but follows from the geometric structure of the boundary value problem. This motivates reformulating the second-order problem as a star-local relaxation update: rather than accessing globally assembled matrices, the algorithm generates the required local boundary and orientation relations directly from the cell index through explicit functions returning the bounding-chain data of local cell stars. We then show how these star-local updates can be compiled into simple and scalable quantum circuits.

We demonstrate the construction for representative div--grad and curl--curl operators. The resulting circuits share a common architecture based on local star information, explicit star-oracles, block encoding of relaxation weights, and interference-based summation. The examples considered here use regularly indexed simplicial and hexahedral complexes, because such indexing makes the star-oracles simple, uniform, and scalable. This regularity concerns primarily the coordinate and indexing structure used for circuit construction; the metric structure enters separately through Hodge-type constitutive maps \cite{Kettunen2014ModelingRotation}. Hence the framework separates the combinatorial data required by the quantum circuit from the metric data describing distances, material parameters, and geometric deformation.

Starting from ordinary scalar-valued differential forms, the same geometric viewpoint also opens a natural route toward vector-valued and covector-valued differential forms, or more generally differential {\em forms with values in vector bundles} \cite{Baez-Muniain,Frankel}. In that setting, the exterior derivative is replaced, or more generally extended, by the covariant exterior derivative. This suggests a path toward quantum algorithms for broader classes of geometrically structured partial differential equations, including problems in continuum mechanics, such as elasticity, as well as geometry processing in computer graphics.

\section{Finite-dimensional operator construction}
\label{sec:finite-dimensional-operator-construction}

We first describe how the continuous second-order operator is represented in a finite-dimensional space. The aim is to start from the local structure of the differential operator itself. The construction begins with the duality between {\em chains} and {\em cochains} \cite{Whitney1957} induced by Stokes' theorem.

\begin{figure}[htbp]
    \centering
    \begin{minipage}[c]{0.30\linewidth}
        \centering
        \includegraphics[width=\linewidth]{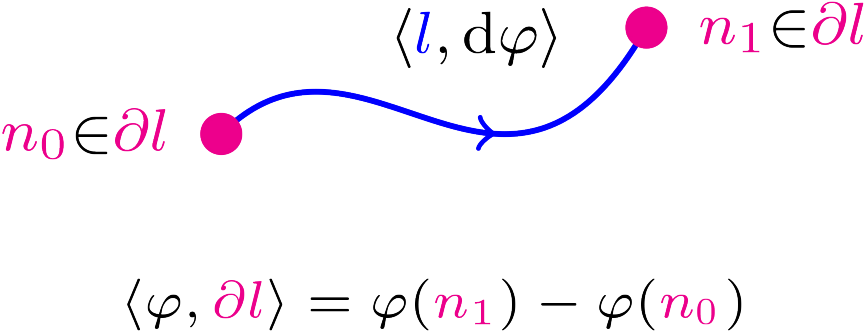}
    \end{minipage}
    \hfill
    \begin{minipage}[c]{0.30\linewidth}
        \centering
        \includegraphics[width=\linewidth]{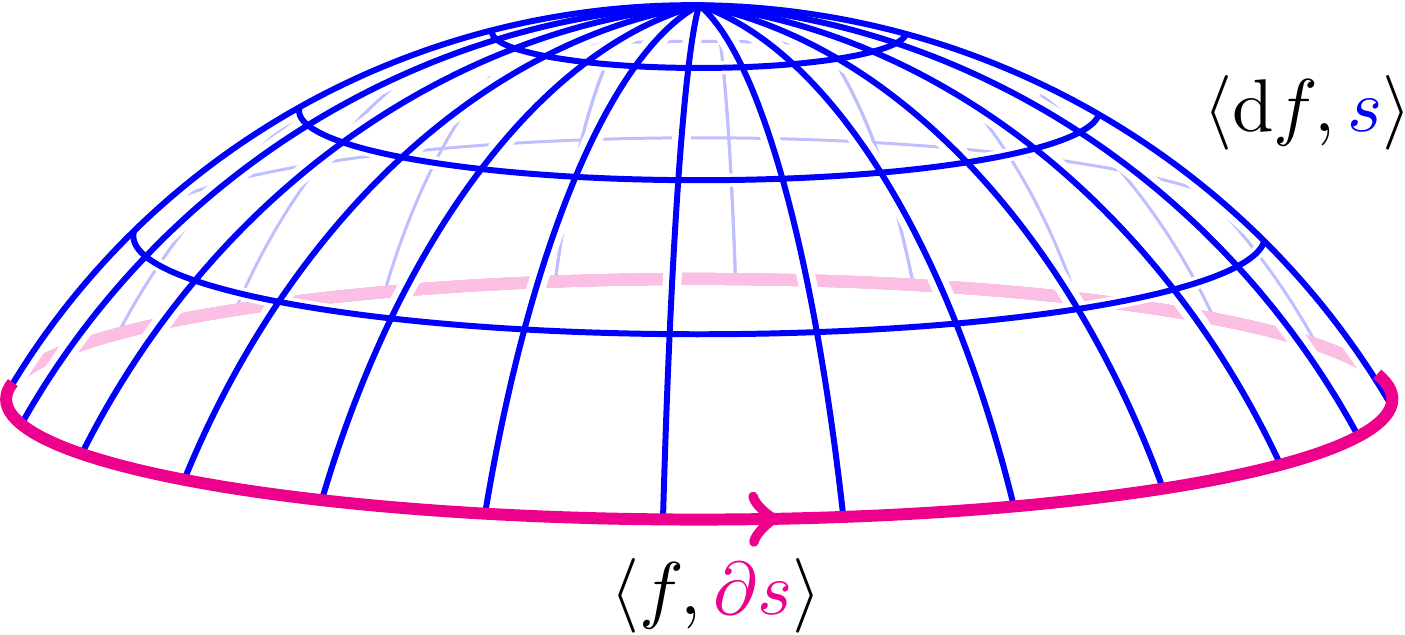}
    \end{minipage}
    \hfill
    \begin{minipage}[c]{0.30\linewidth}
        \centering
        \includegraphics[width=0.85\linewidth]{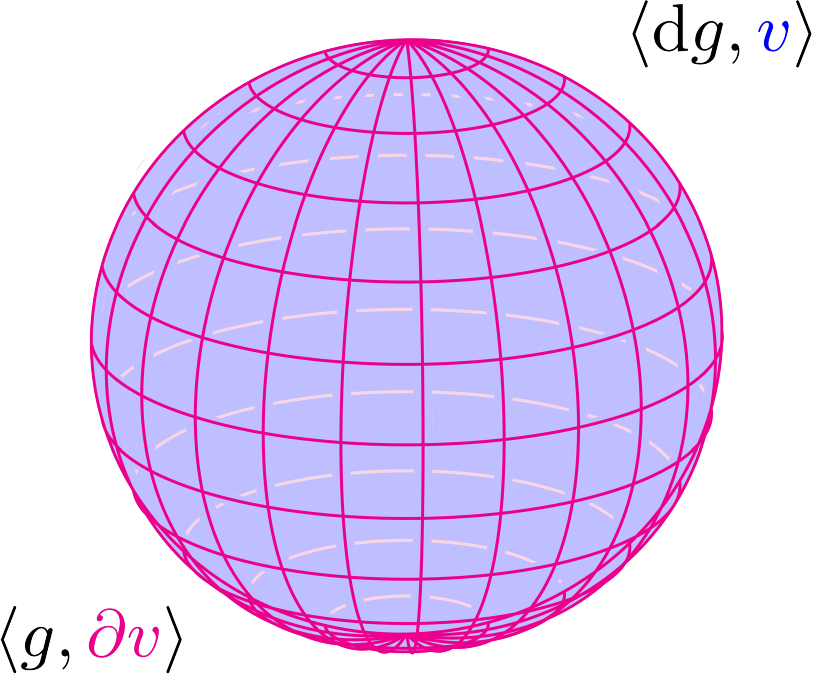}
    \end{minipage}
    \caption{Illustrations of Stokes' theorem in different dimensions. The domain is shown in blue and its boundary in magenta. Left: a curve with boundary nodes. Middle: a surface with boundary curve. Right: a volume with boundary surface. The general Stokes theorem expresses the duality between the exterior derivative and the boundary operator without requiring a metric structure.}
    \label{fig:stokes-theorem}
\end{figure}

Let $\omega$ be an oriented region and let $f$ be a differential form in the domain of the exterior derivative. Stokes' theorem gives
\[
\int_{\omega} {\mathrm d}f = \int_{\partial \omega} f .
\]
Equivalently, using a chain--cochain pairing notation,
\[
\langle {\mathrm d}f,\omega\rangle
=
\langle f,\partial\omega\rangle ,
\]
see figure~\ref{fig:stokes-theorem}.

This identity expresses the duality between the exterior derivative ${\mathrm d}$ on forms and the boundary operator $\partial$ on chains \cite{Whitney1957,Baez-Muniain,Frankel}. It also provides the viewpoint used here. The exterior derivative may be characterized through boundary integrals: requiring ${\mathrm d}f=0$ in a domain is equivalent to requiring
\[
\langle f,\partial\tau\rangle = 0
\]
for all relevant chains $\tau$. Consequently, the pairing also vanishes on the bounding cycles of arbitrarily small, or virtual, neighbourhoods of points.

This point is important for the present construction. On the continuous level, differentiation is not strictly local in the sense of depending only on the value of a field at a single point. Rather, the exterior derivative is characterized by the limiting behaviour of integrals over boundaries of arbitrarily small neighbourhoods. Thus, even though the derivative is evaluated at a point, its definition involves a virtual neighbourhood of that point. By contrast, once the metric and material parameters are fixed, the Hodge-type constitutive relation acts pointwise. Consequently, when passing to a finite-dimensional space, a central question is how to replace the virtual neighbourhoods required by differentiation while preserving the local structure of the continuous operator.

The boundary value problems considered in this work can be written, at the continuous level, as a pair of first-order differential equations together with a constitutive law,
\begin{equation}
\label{eq:first-order-system}
{\mathrm d}f = 0,
\qquad
{\mathrm d}g = u,
\qquad
g = \alpha \star f ,
\end{equation}
where $\star$ denotes the Hodge operator and $\alpha$ represents material or constitutive parameters \cite{Hodge1934,Hodge1941,Frankel}. If $f={\mathrm d}\Lambda$, the system gives a second-order equation of the form
\begin{equation}
\label{eq:continuous-second-order}
{\mathrm d}\star\alpha\, {\mathrm d}\Lambda = u .
\end{equation}
The finite-dimensional task is therefore to construct a counterpart of this continuous second-order problem that preserves the local differential structure of (\ref{eq:first-order-system}).

Assume that the domain $\Omega$ is covered by an oriented cell complex. A $p$-chain is a finite linear combination of oriented $p$-cells \cite{Whitney1957,Munkres1984},
\[
\tau_p = \sum_i \lambda_i \omega_p^i ,
\qquad
\lambda_i \in \mathbb{R}.
\]
The boundary operator extends linearly,
\begin{equation}
\partial \tau_p
=
\sum_i \lambda_i \partial \omega_p^i .
\label{eq:boundary-chain}
\end{equation}
A $p$-cochain assigns a number to each $p$-chain. In the present setting, cochains arise by integration of differential forms over cells,
\[
\langle f,\tau_p\rangle
=
\int_{\tau_p} f .
\]

\begin{figure}[htbp]
    \centering
    \begin{minipage}[c]{0.32\linewidth}
        \centering
        \includegraphics[width=0.8\linewidth]{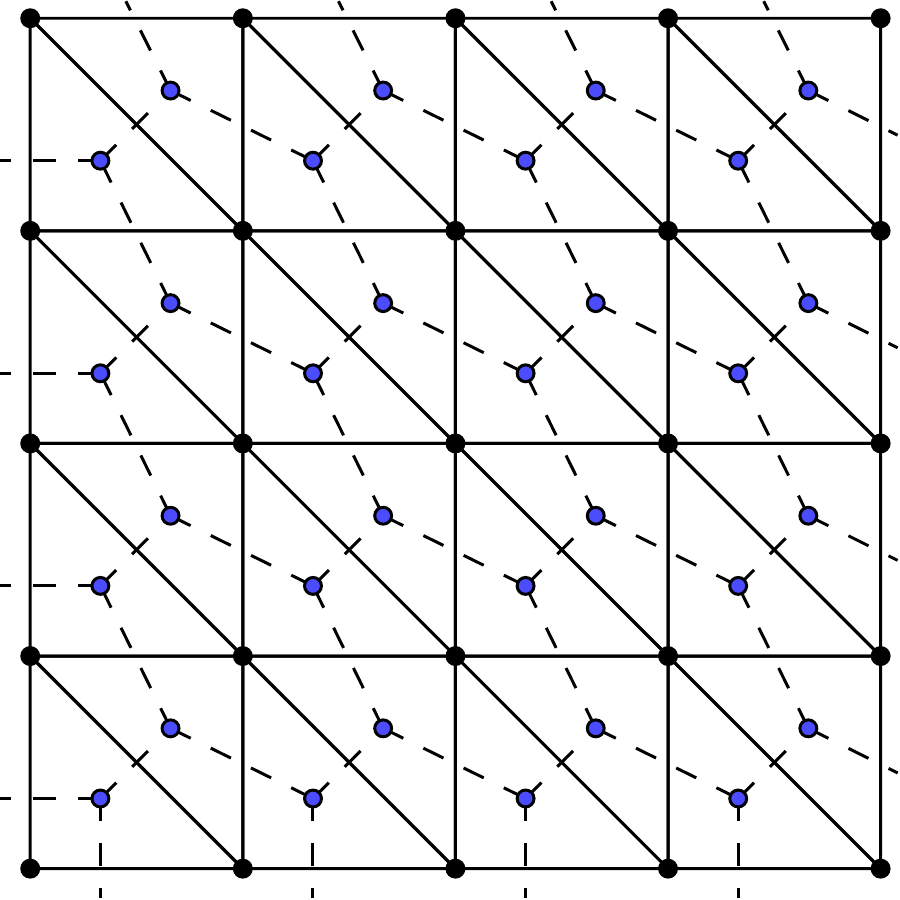}
    \end{minipage}
%    \hspace{0.02\linewidth}
    \begin{minipage}[c]{0.4\linewidth}
        \centering
        \includegraphics[width=0.8\linewidth]{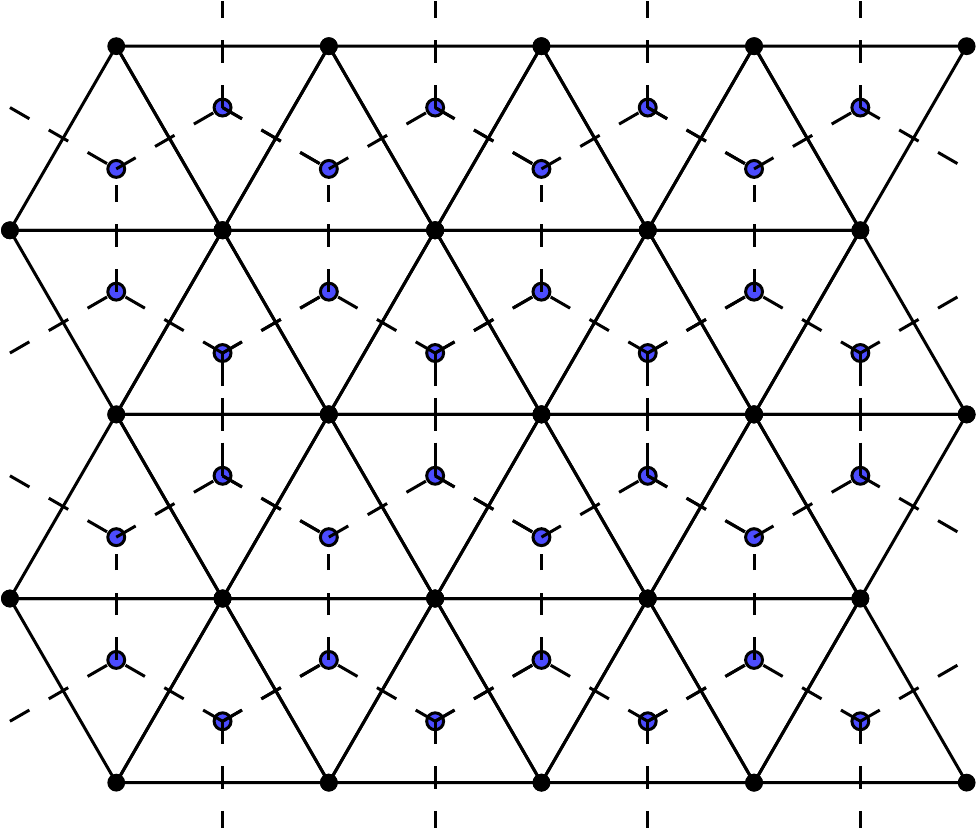}
    \end{minipage}
    \caption{Examples of primal--dual mesh pairs in dimension $n=2$. The primal mesh is shown with solid lines and the dual mesh with dashed lines. A primal $p$-cell is paired with a dual $(n-p)$-cell.}
    \label{fig:dual-mesh}
\end{figure}

A finite-dimensional counterpart of the homogeneous equation ${\mathrm d}f=0$ is obtained by requiring
\begin{equation}
\langle f,\partial \tau_{p+1}\rangle = 0
\label{eq:finite-homogeneous}
\end{equation}
for the relevant bounding chains in the cell complex. Similarly, a finite-dimensional counterpart of ${\mathrm d}g=u$ is obtained by requiring
\begin{equation}
\langle g,\partial \tilde{\tau}_{n-p+1}\rangle
=
\langle u,\tilde{\tau}_{n-p+1}\rangle ,
\label{eq:finite-nonhomogeneous}
\end{equation}
where tildes denote cells and chains in the dual complex; We use a primal--dual cell complex in which each primal $p$-cell $\omega_p^i$ is paired with a dual $(n-p)$-cell $\tilde{\omega}_{n-p}^i$. Such primal--dual constructions are standard in geometric formulations of field problems \cite{Yee1966,Weiland1984,Bossavit-Kettunen1999,Bossavit-book,BossavitPIER2001,Hirani2003}. Examples of primal--dual mesh pairs are shown in figure~\ref{fig:dual-mesh}.

The Hodge-type constitutive relation is represented by a local relation between paired primal and dual cells. Several finite-dimensional realizations of this relation are possible \cite{Tarhasaari-Kettunen-Bossavit1999,Bossavit-Kettunen1999}. One possible choice is the generalized finite-difference form \cite{BossavitPIER2001}, used here for brevity and simplicity, in which this relation reads
\begin{equation}
\frac{\langle g,\tilde{\omega}_{n-p}^i\rangle}
     {|\tilde{\omega}_{n-p}^i|}
=
\alpha_i
\frac{\langle f,\omega_p^i\rangle}
     {|\omega_p^i|}.
\label{eq:finite-hodge}
\end{equation}
Thus the metric and material information is carried by the Hodge-type constitutive factor, while the exterior derivative is represented by oriented boundary relations between cells \cite{Tarhasaari-Kettunen-Bossavit1999,Bossavit-Kettunen1999}.

Assuming trivial topology, the condition ${\mathrm d}f=0$ allows us to write
\[
f = {\mathrm d}\Lambda
\]
for a potential $\Lambda$. (For nontrivial topology, closed forms need not be exact, and additional homological and cohomological constraints are required to select an exact representative and to determine the potential uniquely \cite{Bossavit1998,Suuriniemi2002,Suuriniemi2004,Pellikka2010,Pellikka2013}.) 

In the finite-dimensional setting, the relation $f={\mathrm d}\Lambda$ becomes
\begin{equation}
\langle f,\omega_p^j\rangle
=
\langle {\mathrm d}\Lambda,\omega_p^j\rangle
=
\langle \Lambda,\partial\omega_p^j\rangle
=
\sum_k {\bf D}_{j,k}
\langle \Lambda,\omega_{p-1}^k\rangle ,
\label{eq:finite-exterior-derivative}
\end{equation}
where ${\bf D}$ is the incidence matrix between $p$-cells and $(p-1)$-cells \cite{Bossavit-Kettunen1999,BossavitPIER2001}. Its entries are
\[
{\bf D}_{j,k}
=
\left\{
\begin{array}{ll}
\pm 1, & \hbox{if } \omega_{p-1}^k \subset \partial \omega_p^j,\\[2pt]
0, & \hbox{otherwise},
\end{array}
\right.
\]
with the sign determined by the relative orientations of the cells. In other words, ${\bf D}$ records which lower-dimensional cells bound a higher-dimensional cell, and with which orientation.

\begin{figure}[htbp]
    \centering
    \includegraphics[width=0.25\linewidth]{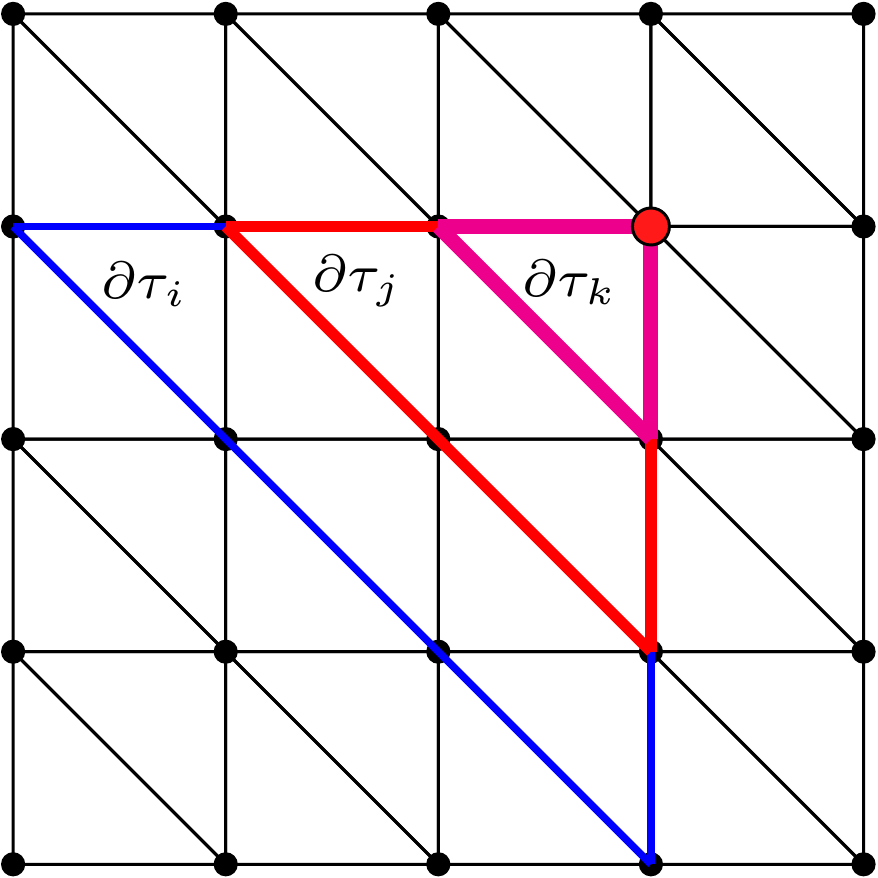}
    \caption{A simplicial complex illustrating bounding cycles at different scales. If $\langle f,\partial\tau\rangle=0$ holds for all relevant chains $\tau$, then the pairing also vanishes on shrinking bounding cycles, shown here in blue, red, and magenta, down to the smallest cycles available in the cell complex. The finite-dimensional construction therefore replaces continuum virtual neighbourhoods by finitely many boundary relations.}
    \label{fig:mesh-boundaries}
\end{figure}

Figure~\ref{fig:mesh-boundaries} illustrates how the finite-dimensional counterpart of ${\mathrm d}f=0$ is imposed through bounding chains in a cell complex.

Substituting (\ref{eq:finite-hodge}) and (\ref{eq:finite-exterior-derivative}) into (\ref{eq:finite-nonhomogeneous}) gives the finite-dimensional counterpart of the second-order problem. For each primal $(p-1)$-cell $\omega_{p-1}^i$, we obtain
\begin{equation}
\sum_j
{\bf D}^{T}_{i,j}
\alpha_j
\frac{|\tilde{\omega}_{n-p}^j|}{|\omega_p^j|}
\sum_k
{\bf D}_{j,k}
\langle \Lambda,\omega_{p-1}^k\rangle
=
\langle u,\tilde{\omega}_{n-p+1}^i\rangle .
\label{eq:finite-second-order-expanded}
\end{equation}
Equivalently,
\begin{equation}
{\bf D}^{T}{\bf H}{\bf D}\,\underline{\Lambda}
=
\underline{u},
\label{eq:finite-second-order-matrix}
\end{equation}
where ${\bf H}$ denotes the finite-dimensional Hodge-type constitutive operator. For the diagonal generalized finite-difference realization introduced above, its entries are
\[
{\bf H}_{j,j}
=
\alpha_j
\frac{|\tilde{\omega}_{n-p}^j|}{|\omega_p^j|}.
\]
The vector $\underline{\Lambda}$ contains the cochain values
$\langle \Lambda,\omega_{p-1}^k\rangle$, while $\underline{u}$ contains the source terms
$\langle u,\tilde{\omega}_{n-p+1}^i\rangle$.

Although (\ref{eq:finite-second-order-matrix}) has the familiar matrix form, the matrix is not the primary object in the present construction. The nonzero entries of ${\bf D}$ and ${\bf D}^{T}$ only encode local boundary and orientation relations. 

To obtain a matrix-free update rule, we isolate the self-contribution in (\ref{eq:finite-second-order-expanded}). Writing
\[
\Lambda_i
=
\langle \Lambda,\omega_{p-1}^i\rangle,
\qquad
u_i
=
\langle u,\tilde{\omega}_{n-p+1}^i\rangle,
\]
we obtain
\begin{equation}
\Lambda_i
=
\frac{1}{\Delta_i}
\left(
\sum_j
{\bf D}^{T}_{i,j}{\bf H}_{j,j}
\sum_{k\neq i}
{\bf D}_{j,k}\Lambda_k
-
u_i
\right),
\label{eq:fixed-point-form}
\end{equation}
where
\begin{equation}
\Delta_i
=
-
\sum_j
{\bf D}^{T}_{i,j}{\bf H}_{j,j}{\bf D}_{j,i}.
\label{eq:Delta-definition}
\end{equation}
Introducing a relaxation parameter $0<\beta<1$ gives the iterative update
\begin{equation}
\Lambda_i^{s+1}
=
\frac{\beta}{\Delta_i}
\left(
\sum_j
{\bf D}^{T}_{i,j}{\bf H}_{j,j}
\sum_{k\neq i}
{\bf D}_{j,k}\Lambda_k^s
-
u_i
\right)
+
(1-\beta)\Lambda_i^s .
\label{eq:star-local-update}
\end{equation}
The iteration is continued until, for example,
\[
|\underline{\Lambda}^{s+1}-\underline{\Lambda}^{s}|
<
\mathrm{tolerance}.
\]
Equation~(\ref{eq:star-local-update}) is the central finite-dimensional update rule used in this work. Its support is determined by those cells that are connected to $\omega_{p-1}^i$ through the boundary relations encoded by ${\bf D}$ and ${\bf D}^{T}$. In other words, only the cells in the star of $\omega_{p-1}^i$ contribute to the update of $\Lambda_i$. Thus the star appears as the finite-dimensional counterpart of the virtual neighbourhood required by differentiation in the continuous definition of the operator.

The update also exposes the structural separation underlying the quantum construction. The matrices ${\bf D}$ and ${\bf D}^{T}$ encode the differential, or combinatorial, structure through local boundary and orientation relations, whereas ${\bf H}$ carries the metric and material data through the Hodge-type constitutive relation. In the quantum implementation, the former are generated by explicit star-oracles from the cell index, while the latter enter through the block-encoding of scalar weights.

\subsection{Examples: div--grad and curl--curl operators}
\label{sec:finite-dimensional-examples}

We next spell out two representative cases. These examples also fix the local update patterns that are later compiled into quantum circuits.

\subsubsection{Operator div--grad}

\begin{figure}[htbp]
    \centering
    \includegraphics[width=0.3\linewidth]{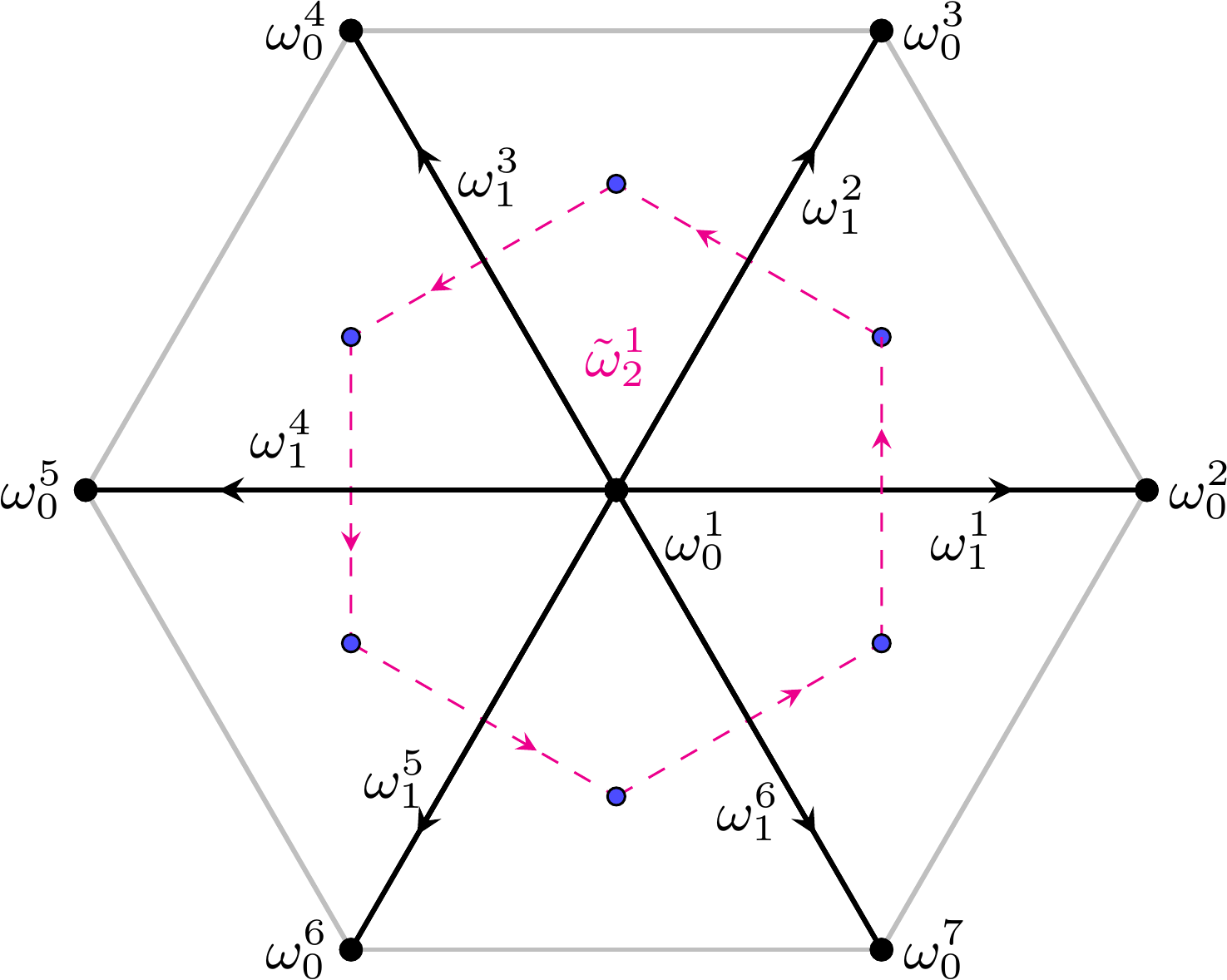}
    \caption{The local neighbourhood, or star, of node $\omega_0^1$ in a complex of regular triangles, together with the dual $2$-cell $\tilde{\omega}_2^1$ paired with $\omega_0^1$.}
    \label{fig:laplace-star}
\end{figure}

In dimension $n=2$, the div--grad operator is the metric counterpart of operator ${\mathrm d}\star{\mathrm d}$ acting on $0$-forms. We write the unknown as $\varphi$ to emphasize that it is a scalar potential. On a regular triangular complex, the star of an interior node contains six neighbouring nodes; see figure~\ref{fig:laplace-star}.

For one such node, the local incidence pattern may be written as
\begin{equation}
{\bf D}
=
\setlength\arraycolsep{4pt}
\begin{bmatrix}
-1 & 1 & 0 & 0 & 0 & 0 & 0 \\
-1 & 0 & 1 & 0 & 0 & 0 & 0 \\
-1 & 0 & 0 & 1 & 0 & 0 & 0 \\
-1 & 0 & 0 & 0 & 1 & 0 & 0 \\
-1 & 0 & 0 & 0 & 0 & 1 & 0 \\
-1 & 0 & 0 & 0 & 0 & 0 & 1
\end{bmatrix}.
\label{eq:divgrad-local-D}
\end{equation}
For a constant ratio
\[
{\bf H}_{j,j}
=
\alpha
\frac{|\tilde{\omega}_1^j|}{|\omega_1^j|}
\]
over the six incident edges, the update (\ref{eq:star-local-update}) becomes
\begin{equation}
\varphi_1^{s+1}
=
\frac{\beta}{6}
\left(
\varphi_2^s+\varphi_3^s+\cdots+\varphi_7^s
-
\frac{|\omega_1^1|}{|\tilde{\omega}_1^1|}u_1
\right)
+
(1-\beta)\varphi_1^s .
\label{eq:divgrad-update}
\end{equation}
Thus the update of the scalar value at the central node uses only the values associated with its local star. This is the finite-dimensional counterpart of imposing the Poisson-type equation
\[
{\mathrm{div}}\,{\rm grad}\,\varphi = u
\]
in a local neighbourhood. The approach solving the Poisson problem with \eqref{eq:divgrad-update} coincides with the classical Lattice Boltzmann Method \cite{Krueger2016, Mohammed2019}.

\subsubsection{Operator curl--curl}

\begin{figure}[htbp]
    \centering
    \includegraphics[width=0.3\linewidth]{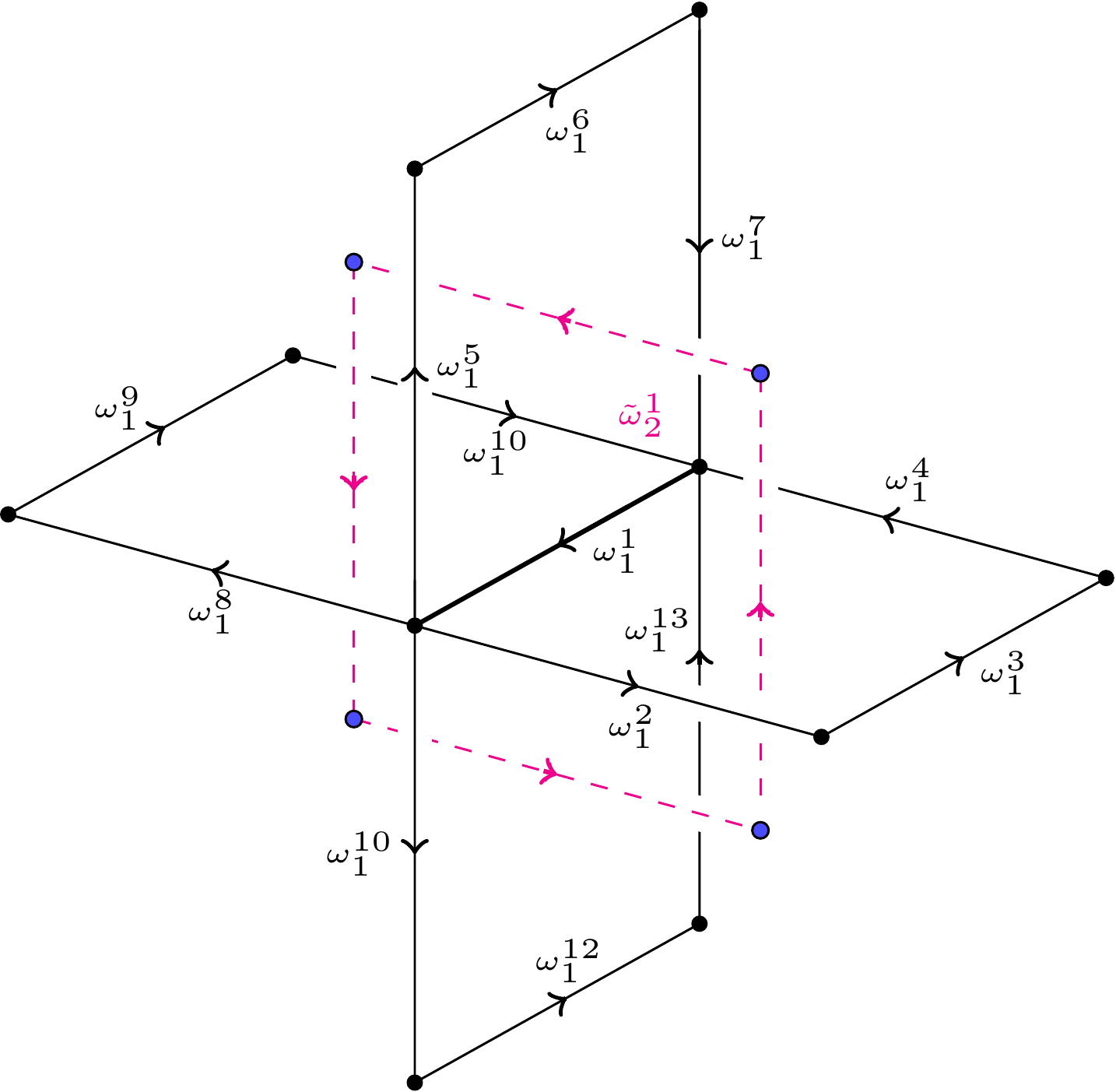}
    \caption{The star of edge $\omega_1^1$ in a regular hexahedral complex, together with the dual facet $\tilde{\omega}_2^1$ paired with $\omega_1^1$.}
    \label{fig:curlcurl-star}
\end{figure}

In dimension $n=3$, the operator ${\mathrm d}\star{\mathrm d}$ acting on $1$-forms corresponds to the curl--curl operator. The unknown is now a $1$-cochain, denoted by $a$, associated with oriented edges. On a regular hexahedral complex, the star of an interior edge contains four incident faces, and the corresponding local update involves twelve neighbouring edge values; see figure~\ref{fig:curlcurl-star}.

A local incidence pattern for the star of an edge can be written as
\begin{equation}
{\bf D}
=
\setlength\arraycolsep{4pt}
\begin{bmatrix}
1 & 1 & 1 & 1 & 0 & 0 & 0 & 0 & 0 & 0 & 0 & 0 & 0\\
1 & 0 & 0 & 0 & 1 & 1 & 1 & 0 & 0 & 0 & 0 & 0 & 0\\
1 & 0 & 0 & 0 & 0 & 0 & 0 & 1 & 1 & 1 & 0 & 0 & 0\\
1 & 0 & 0 & 0 & 0 & 0 & 0 & 0 & 0 & 0 & 1 & 1 & 1
\end{bmatrix}.
\label{eq:curlcurl-local-D}
\end{equation}
For the corresponding problem
\[
{\rm curl}\,{\rm curl}\,a = j,
\]
the local relaxation update takes the form
\begin{equation}
a_1^{s+1}
=
-\frac{\beta}{4}
\left(
a_2^s+a_3^s+\cdots+a_{13}^s
-
\frac{|\omega_2^1|}{|\tilde{\omega}_1^1|}j_1
\right)
+
(1-\beta)a_1^s .
\label{eq:curlcurl-update}
\end{equation}
If needed, the uniqueness of $a$---that is, the so-called gauge condition---can be imposed with the spanning tree extraction method \cite{AlbaneseRubinacci1988, Suuriniemi2002}. Again, the update is supported only on the star of the edge. The role of the finite-dimensional construction is therefore to replace the virtual neighbourhoods of the continuous second-order operator by explicitly computable cell stars.

The two examples are instances of the same finite-dimensional operator construction. In both cases, the exterior derivative is represented by oriented boundary relations, the Hodge-type constitutive operator carries the metric and material information, and the resulting second-order operator acts on the star of the cell carrying the unknown. The star-local update is therefore the finite-dimensional form of the construction derived above, and it is this object that is compiled into quantum circuits in the following sections.

\section{Star-local quantum circuit construction}
\label{sec:star-local-quantum-circuit}

The finite-dimensional construction in section~\ref{sec:finite-dimensional-operator-construction} shows that one relaxation step of the second-order problem can be written as a star-local update. The quantum-circuit task is therefore not to implement a globally assembled matrix ${\bf D}^{T}{\bf H}{\bf D}$, but to implement the local operations appearing in the update rule (\ref{eq:star-local-update}). In this section we describe the corresponding quantum-circuit architecture.

One circuit execution implements one relaxation step. The current iterate is encoded as a quantum state, local star data are generated from the cell index by a star-oracle, scalar relaxation weights are applied by block encoding, and the required contributions are combined by interference to produce the next iterate. We present the construction for the curl--curl update; the same architecture applies to the div--grad update with the corresponding change of cell degree and local star structure.

\subsection{Star-oracle as a replacement for incidence-matrix access}
\label{sec:star-oracle}

The update rule (\ref{eq:star-local-update}) depends only on the local boundary and orientation relations associated with the star of the cell carrying the degree of freedom. For the curl--curl case, the degrees of freedom are carried by oriented primal edges. We therefore require a map that, given an oriented edge $e$, returns the signed neighbouring edge data needed in the local update.

We denote this map by
\[
\mathcal{S}:\; e \longmapsto \mathrm{Star}_{1}(e),
\]
where $\mathrm{Star}_{1}(e)$ is the signed $1$-chain obtained from the bounding-chain data of the star of $e$. The role of $\mathcal{S}$ is to replace access to stored rows of an incidence matrix by a function call that generates the required local information directly from the edge index.

To define $\mathrm{Star}_{1}(e)$, let $\mathrm{Star}(e)$ denote the set of oriented facets $f$ incident to $e$. The incidence coefficient
\[
[\partial f,e]\in\{-1,+1\}
\]
records whether the orientation induced on $e$ by the boundary $\partial f$ agrees with the chosen orientation of $e$. We define
\[
\lambda_f=[\partial f,e],
\qquad
\hat f(e)=\lambda_f f,
\]
so that the coefficient of $e$ in $\partial\hat f(e)$ is $+1$ for every incident facet. Define
\[
F^\star(e)
=
\sum_{f\in \mathrm{Star}(e)} \hat f(e)
\]
and
\begin{equation}
\mathrm{Star}_{1}(e)
=
\partial F^\star(e)
-
|\mathrm{Star}(e)|\,e .
\label{eq:star1-def-circuit}
\end{equation}
Here $|\mathrm{Star}(e)|$ denotes the number of oriented facets incident to $e$. Since the orientation of each $\hat f(e)$ has been chosen so that $e$ appears in $\partial\hat f(e)$ with coefficient $+1$, the central edge occurs exactly $|\mathrm{Star}(e)|$ times in $\partial F^\star(e)$. The subtraction in (\ref{eq:star1-def-circuit}) removes these repeated central-edge contributions. The remaining signed $1$-chain contains precisely the neighbouring edge data entering the curl--curl update.

For a general unstructured complex, evaluating $e\mapsto \mathrm{Star}_{1}(e)$ would require stored incidence data. In the curl--curl construction used here, we instead assume a regularly indexed hexahedral cell complex with cyclic symmetry between the coordinate directions. Edges parallel to the $x$-, $y$-, and $z$-axes are indexed separately, and each directional edge family is embedded into an index set of size $n=2^m$ so that it can be addressed by an $m$-qubit index register. If the number of edges in a given family is smaller than $2^m$, the remaining indices are treated as dummy edges with zero cochain values.

With this indexing, the same combinatorial star data can be generated by closed-form index transformations. The regularity used here concerns the index and incidence structure only; metric and material information enters separately through the Hodge-type operator ${\bf H}$. The complete indexing convention, including the treatment of dummy indices, is given in appendix~\ref{app:indexing-symmetry}. In the quantum implementation, the local boundary and orientation relations are generated on the fly.

For example, consider an interior edge of the regular hexahedral complex oriented in the positive $x$-direction. We denote this edge by
$\omega_{1x}^{(i,j,k)}$, where the subscript $1x$ indicates a primal $1$-cell parallel to the $x$-axis and the superscript $(i,j,k)$ gives the index of its initial node, so that
$\omega_{1x}^{(i,j,k)}$ is oriented from $\omega_0^{(i,j,k)}$ to $\omega_0^{(i+1,j,k)}$; see figure~\ref{fig:indexing_of_cells} in appendix~\ref{app:indexing-symmetry}. For this edge, the oracle returns the signed neighbouring edge chain
\begin{equation}
\begin{alignedat}{4}
\mathrm{Star}_{1}\!\bigl(\omega_{1x}^{(i,j,k)}\bigr)
={}&{}&
-&\, \omega_{1z}^{(i,j,k)}
&{}-&\, \omega_{1x}^{(i,j,k+1)}
&{}+&\, \omega_{1z}^{(i+1,j,k)}
\\
&{}&
+&\, \omega_{1z}^{(i,j,k-1)}
&{}-&\, \omega_{1z}^{(i+1,j,k-1)}
&{}-&\, \omega_{1x}^{(i,j,k-1)}
\\
&{}&
+&\, \omega_{1y}^{(i+1,j,k)}
&{}-&\, \omega_{1x}^{(i,j+1,k)}
&{}-&\, \omega_{1y}^{(i,j,k)}
\\
&{}&
-&\, \omega_{1x}^{(i,j-1,k)}
&{}-&\, \omega_{1y}^{(i+1,j-1,k)}
&{}+&\, \omega_{1y}^{(i,j-1,k)} .
\end{alignedat}
\label{eq:main-star1-xedge}
\end{equation}
Here $\omega_{1y}^{(i,j,k)}$ and $\omega_{1z}^{(i,j,k)}$ denote analogous oriented primal edges parallel to the $y$- and $z$-axes. Thus equation~(\ref{eq:main-star1-xedge}) gives the data produced by the star-oracle: a fixed pattern of shifted neighbouring edge indices together with their orientation signs. In the circuit, the shifts are implemented by controlled index transformations and the signs by controlled phase operations. The rules for $y$- and $z$-directed central edges are obtained by cyclic relabelling of the coordinate directions. The full boundary-aware formula, including admissibility predicates for boundary edges, together with the complete indexing convention and dummy-index treatment, is given in appendix~\ref{app:indexing-symmetry}.

\subsection{Regular indexing and cyclic symmetry}
\label{sec:regular-indexing-cyclic-symmetry}

In the curl--curl construction, the unknown is a $1$-cochain on oriented primal edges. We use an indexing in which edges parallel to the $x$-, $y$-, and $z$-directions are grouped separately and oriented along the positive coordinate axes. This indexing makes the cyclic symmetry of the indexed cell complex, and hence of the star-oracle, explicit. A local rule derived for one coordinate direction can then be transferred to the other two directions by cyclic relabelling.

The combination of regular indexing, cyclic symmetry, and a star-oracle is naturally compatible with quantum-walk primitives \cite{Venegas_Andraca_2012,Shakeel_2020}. The action of the star-oracle on the index register can be implemented by controlled basis-state shifts: the oracle maps the index of an edge to the indices of neighbouring edges in its local star, while additional controlled phase operations (P) encode the corresponding orientation signs. Accordingly, the required index updates are implemented using left(L) and right(R) shift operators of the type used in \cite{Budinski2021ADE}, which are shown in~\ref{qc:shift operators}. When circuit depth is the primary concern, these shifts may also be replaced by their parallelised variants \cite{budinski2023efficient}.

Let $C_{1x}$, $C_{1y}$, and $C_{1z}$ denote the sets of oriented edges parallel to the three coordinate axes, and let $A_x$, $A_y$, and $A_z$ denote the corresponding vectors of $1$-cochain values. Similarly, let $J_x$, $J_y$, and $J_z$ denote the source terms on dual $2$-cells paired with these edges. The cyclic permutation $\pi$ acts by
\begin{equation}
\pi\cdot(x,y,z;x)=(y,z,x;y),
\qquad
\pi^2\cdot(x,y,z;x)=(z,x,y;z).
\label{eq:cyclic-symmetry}
\end{equation}
Thus, a rule for the star-oracle derived for one coordinate direction can be transferred to the other two directions by relabelling.

To exploit this symmetry, we use the packed iterate vector
\begin{equation}
\Lambda^s =
[A_x,A_y,A_z,J_z,J_y,J_x]^T .
\label{eq:packed-iterate}
\end{equation}
The reversed ordering of the source components is a bookkeeping convention chosen to align the source branches with the corresponding edge-component branches in the circuit.

Let $n=2^m$ denote the number of indexed cochain values in each component block. The packed vector has dimension $6n$. An index register of $m=\log_2 n$ qubits addresses the entries inside each component block. In addition, three label qubits $\lambda_p$ address the packed component blocks, while two label qubits $\lambda_d$ address the component direction of the edge being updated. The circuit also uses a three-qubit selector register $d$ and two ancilla qubits $a$. Thus the total number of qubits is
\begin{equation}
\begin{split}
Q
&=
m+\lambda_p+\lambda_d+d+a
\\
&=
\log_2 n + 3 + 2 + 3 + 2 .
\end{split}
\label{eq:qubit-count}
\end{equation}
The index-register cost therefore grows logarithmically with the number of indexed degrees of freedom. The remaining overhead is constant for the present curl--curl circuit architecture: five qubits label the packed components and edge direction, three qubits select the computational branch, and two ancilla qubits are used for controlled weighting operations and for isolating the source terms $J$ in connection with dynamic circuit timestepping \cite{bastida2026quantum}. The price of this uniform register design is the cyclic packing in (\ref{eq:packed-iterate}) and the use of dummy indices introduced by padding. These components simplify the oracle and make the circuit structure independent of the spatial location and coordinate direction of the edge.

\subsection{Register structure}
\label{sec:register-structure}

The circuit acts on four types of registers:
\begin{itemize}
    \item an index register $\ket{e}_m$ encoding the oriented edge index;
    \item a component-label register $\ket{\ell}_{\lambda_p, \lambda_d}$ identifying the component of the packed iterate vector;
    \item a selector register $\ket{k}_d$ controlling the computational branch, including star contributions, source terms, and retained components;
    \item an ancilla qubits $\ket{q}_a$ used for controlled operations and block encoding.
\end{itemize}

The input state $\ket{\Lambda^s}$ is prepared by a state-preparation routine, treated here as a black box \cite{Shende_quantum_state,PhysRevA.93.032318}. Cyclic controlled permutations ($\pi$) are applied on the subspaces created using $H$ gates, following some rearrangement of order of components making way for the application of star-oracle created in eq.~\ref{eq:main-star1-xedge}.  The star-oracle acts coherently on the index register, conditioned on the selector and component-label registers. Its purpose is to route amplitudes from an edge index to the neighbouring edge indices appearing in $\mathrm{Star}_{1}(e)$.

\subsection{One-step quantum relaxation algorithm}
\label{sec:one-step-quantum-relaxation}

The circuit blocks shown in Fig.~\ref{fig:qc-full} implements one relaxation step of the star-local update. At the algorithmic level, one step consists of the following operations:
\begin{enumerate}
    \item Prepare the packed iterate state $\ket{\Lambda^s}$, together with the component-label, selector, and ancilla registers.

    \item Apply Hadamard gates to the selector register to create computational branches, including a designated branch containing a scaled copy of $\ket{\Lambda^s}$ for dynamic circuit timestepping.

    \item Apply the controlled cyclic permutations $\pi$ and $\pi^2$ on the appropriate computational branches.

    \item Apply the star-oracle $\mathcal{S}$ to generate, coherently and from the cell index, the neighbouring cell indices and orientation signs required by the local update.

    \item Apply the block-encoding stage to insert the scalar coefficients appearing in the relaxation rule, including relaxation weights, retained-term weights, source-term contributions, and Hodge-type constitutive factors.

    \item Align computational branches corresponding to the same output index using reversible operations such as controlled swaps and CNOT gates.

    \item Apply Hadamard interference to combine the routed and weighted contributions into the designated output subspace.

    \item Decode the output subspace to obtain the next iterate $\Lambda^{s+1}$, impose boundary conditions and constraints classically, and reinitialise the state for the next relaxation step.
\end{enumerate}

Thus, one execution of the circuit implements the relaxation map
\[
\Lambda^s \longmapsto \Lambda^{s+1}
\]
at the level of the designated output amplitudes. Equivalently, the circuit maps the prepared input state to a state whose selected output subspace contains $\Lambda^{s+1}$ up to known normalization, block-encoding, and interference factors. The circuit provides all the components required for the dynamic circuit timestepping introduced in~\cite{bastida2026quantum}. 

\subsection{Block encoding of relaxation weights}
\label{sec:block-encoding-relaxation-weights}

After the star-oracle has generated the required local contributions, the scalar coefficients in the relaxation update must be applied. For the curl--curl update (\ref{eq:curlcurl-update}), these include the relaxation coefficient $\beta/4$, orientation signs, source-term signs, and the retained-term coefficient $1-\beta$.

Direct multiplication of amplitudes by arbitrary scalar coefficients is not unitary in general. We therefore embed the required diagonal or block-diagonal weighting into a larger unitary using block encoding \cite{camps2022fable,camps2023explicit}. Operationally, the required scalar weights are encoded by ancilla-controlled single-qubit $Y$ rotations. The desired weighted terms then appear in the selected block or postselected computational subspace of the enlarged unitary transformation.

This step separates the numerical weighting part of the update from the combinatorial star-oracle. The star-oracle generates which neighbouring values are needed, whereas the block-encoding stage applies the relaxation and constitutive weights associated with the update.

\subsection{Summation by interference}
\label{sec:summation-by-interference}

After the star-oracle and block-encoding stages, the terms entering the local update are distributed across several computational subspaces. To form the next iterate, contributions associated with the same edge index must be added coherently.

This is achieved by first aligning the relevant computational branches so that amplitudes to be added occupy matching basis states. Reversible operations such as controlled swaps and CNOT gates are used for this alignment. Hadamard interference is then used to form element-wise sums and differences.

For example, for a two-qubit state
\[
\ket{\Psi}
=
\alpha\ket{00}
+
\beta\ket{01}
+
\gamma\ket{10}
+
\lambda\ket{11},
\]
applying a Hadamard gate to the first qubit gives
\[
(H\otimes I)\ket{\Psi}
=
\frac{1}{\sqrt{2}}
\left[
(\alpha+\gamma)\ket{00}
+
(\beta+\lambda)\ket{01}
+
(\alpha-\gamma)\ket{10}
+
(\beta-\lambda)\ket{11}
\right].
\]
Thus amplitudes associated with matching indices interfere constructively or destructively. In the full circuit, this mechanism is used to combine the star contributions, source terms, and retained components into the designated output subspace containing the updated iterate.

\begin{figure}[H]
  \centering
  \includegraphics[width=0.6\linewidth]{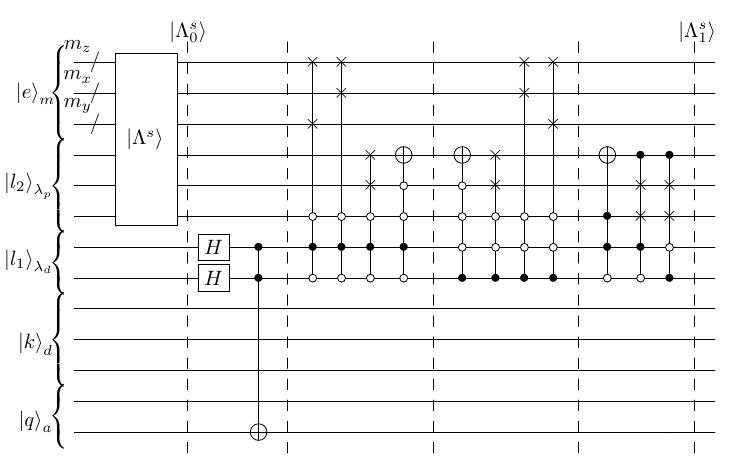}
  \includegraphics[width=0.7\linewidth]{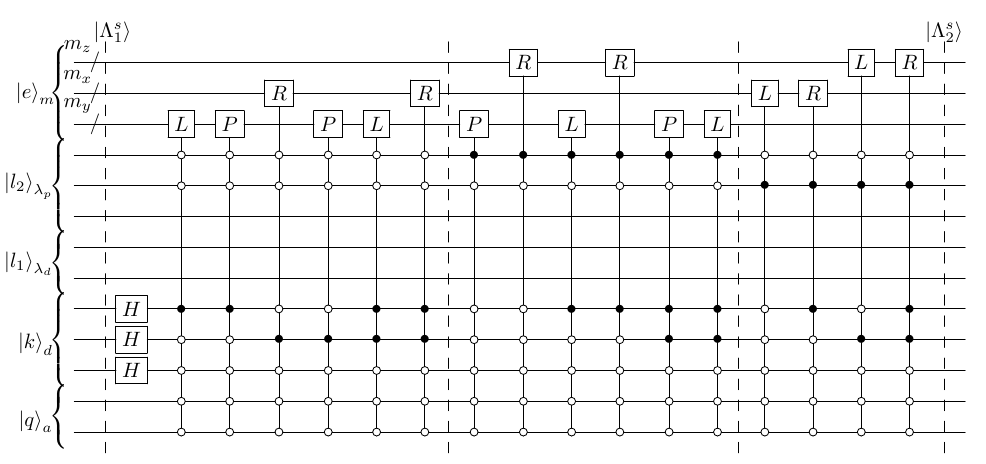}
  % \vspace{0.1em}
  \includegraphics[width=0.6\linewidth]{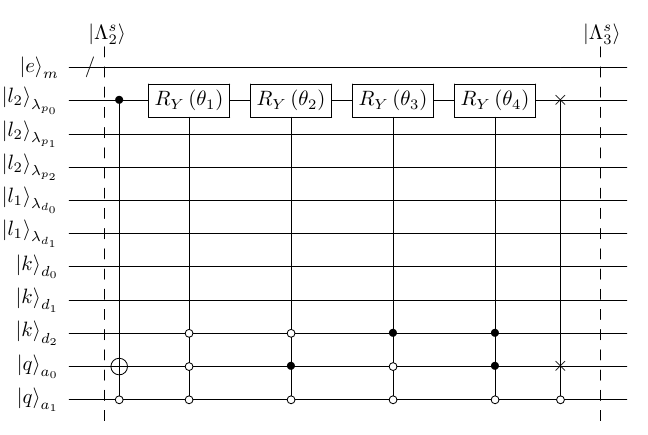}
   % \vspace{0.1em}
  \includegraphics[width=0.7\linewidth]{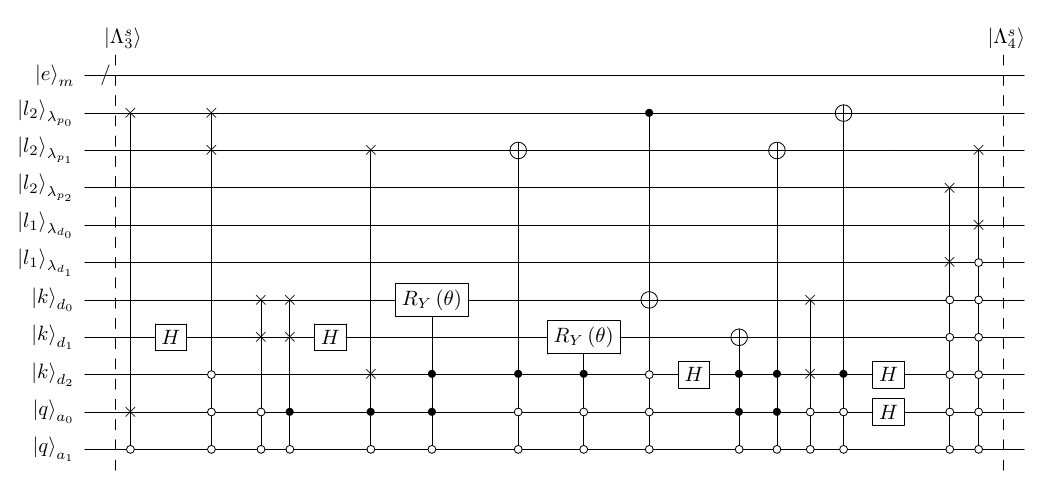}
\caption{Overview of one curl--curl relaxation step. Top: state preparation, register routing and cyclic symmetry, second: star-oracle, quantum circuits of operators \textit{R,L} and \textit{P} are shown in ~\ref{qc:shift operators},~\ref{qc:p}. Third: block encoding of the relaxation and retained-term weights. Bottom: interference-based summation producing the next iterate. The panels are shown stacked for readability; the full circuit is obtained by concatenating them.}
  \label{fig:qc-full}
\end{figure}

\subsection{Circuit architecture for one curl--curl relaxation step}
\label{sec:curlcurl-circuit-architecture}

The circuit for one curl--curl relaxation step consists of four stages:
\begin{itemize}
    \item cyclic permutations ($\pi$) and ($\pi^2$), which generates required structures of variables in subspaces to carry our update for the three vector components.
    \item star-oracle routing, which generates the signed local star contributions from the edge index;
    \item block encoding, which applies the relaxation and retained-term coefficients in a unitary manner;
    \item interference-based summation, which combines the weighted contributions into the next iterate.
\end{itemize}

These four stages are shown schematically in figure~\ref{fig:qc-full}. The figure is displayed as stacked panels for readability, but the full circuit is obtained by concatenating the stages.

At the state level, the four panels of figure~\ref{fig:qc-full} correspond to the sequence
\[
\ket{\Lambda_0^s}
\longrightarrow
\ket{\Lambda_1^s}
\longrightarrow
\ket{\Lambda_2^s}
\longrightarrow
\ket{\Lambda_3^s}
\longrightarrow
\ket{\Lambda_4^s}.
\]
Here $\ket{\Lambda_0^s}$ denotes the initialized and branched input state,$\ket{\Lambda_1^s}$ the state after cyclic permutations ($\pi$) and ($\pi^2$), $\ket{\Lambda_2^s}$ the state after star-oracle routing, $\ket{\Lambda_3^s}$ the state after block encoding of the scalar weights, and $\ket{\Lambda_4^s}$ the state after interference-based summation. The desired next iterate $\Lambda^{s+1}$ is contained in a designated output subspace of $\ket{\Lambda_4^s}$. 

\begin{figure}[htbp]
    \centering
    \includegraphics[width=0.65\linewidth]{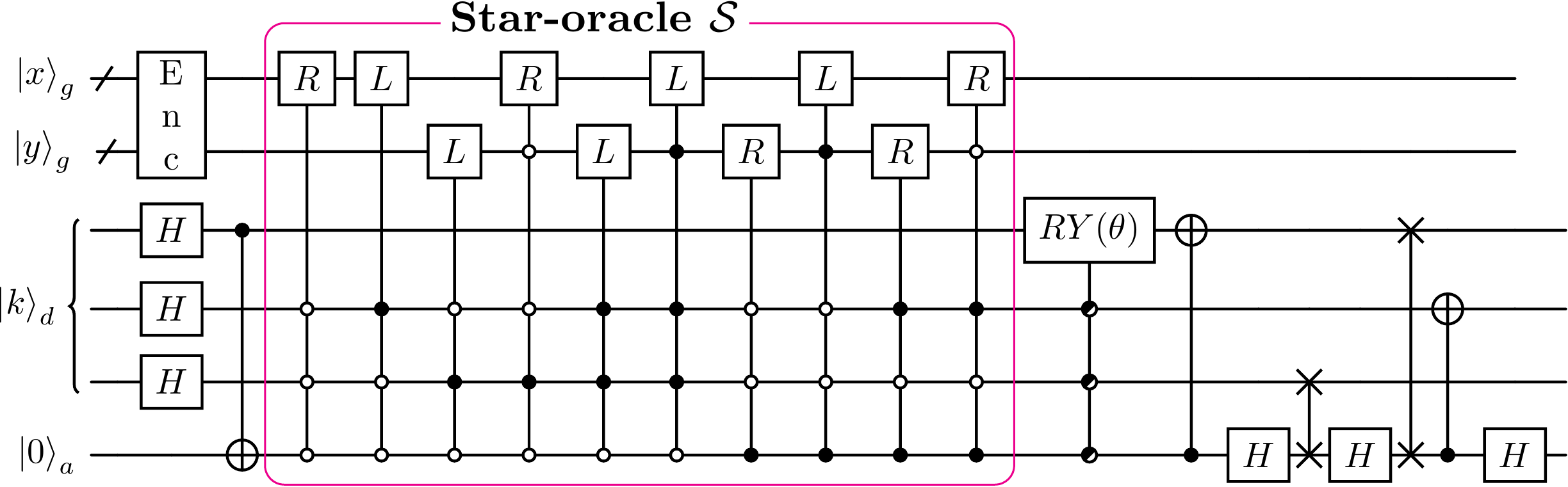}
    \caption{Overview of the div--grad relaxation-step circuit on a regular simplicial complex. The circuit has the same star-local structure as the curl--curl construction, but the local oracle acts on node stars rather than edge stars.}
    \label{fig:qc-divgrad}
\end{figure}

The same star-local architecture also applies to the div--grad update, with the star-oracle changed from stars of oriented edges to stars of nodes. The corresponding div--grad circuit is shown in figure~\ref{fig:qc-divgrad}.

\begin{figure}[htb!]
    \centering
    % \hspace{-1.25cm}
    \includegraphics[width=\textwidth]{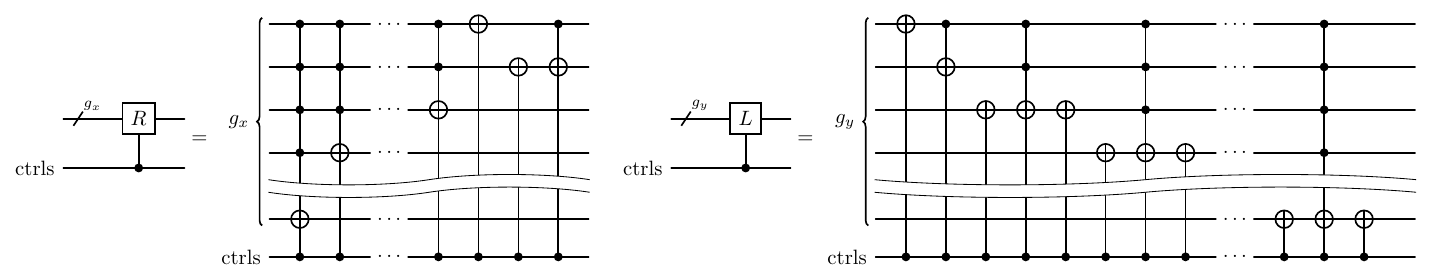}
    \caption{Quantum circuits for \textit{R} and \textit{L} shift operators. Controls are as specified in fig.~\ref{fig:qc-full}}
    \label{qc:shift operators}
\end{figure}

\begin{figure}[htb!]
    \centering
    \includegraphics[]{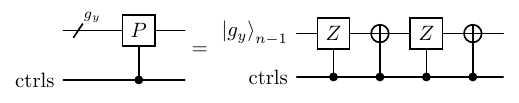}
    \caption{\(P\) operator creating negative amplitudes for terms in eq.~\ref{eq:main-star1-xedge}}
    \label{qc:p}
\end{figure}

\subsection{Iteration protocol and readout}
\label{sec:iteration-protocol-readout}

The complete solver uses an encode--evolve--measure--reinitialise loop. One circuit execution produces one relaxation update, after which boundary conditions, gauge constraints, dummy-index conditions, and stopping criteria are imposed classically. The resulting vector is then re-encoded for the next iteration.

In state-vector simulation, the amplitudes in the designated output subspace are directly accessible. After applying the known normalization factors and accounting for Hadamard prefactors introduced during the summation stage, these amplitudes yield the entries of the next iterate.

On hardware, computational-basis measurements return probabilities and therefore do not directly reveal the signs or phases of amplitudes. In the regularly indexed hexahedral setting considered here, this can be mitigated by adding constant directional offsets to the encoded $1$-cochain values. Since a constant $1$-cochain has vanishing exterior derivative, these offsets do not change ${\mathrm d}a$. If the offsets are chosen sufficiently large, the relevant encoded values can be shifted to a non-negative range, avoiding explicit sign recovery in the targeted output subspace. Further readout details are given in appendix~\ref{app:measurement_readout}.

The present construction focuses on the circuit implementation of one star-local relaxation step. It does not optimize state preparation, readout, or convergence acceleration. These components are treated as part of the outer encode--evolve--measure--reinitialise loop and are discussed in section~\ref{sec:scope-limitations} and appendix~\ref{app:measurement_readout}.

\section{Results and discussion}
\label{sec:results-discussion}

The main result of this work is the construction of a systematic route from continuous second-order differential operators to quantum circuits. The construction proceeds in two steps. First, the continuous operator is represented in a finite-dimensional cochain space as a star-local relaxation update, as described in section~\ref{sec:finite-dimensional-operator-construction}. Second, this star-local update is compiled into a quantum circuit using the architecture described in section~\ref{sec:star-local-quantum-circuit}. The numerical examples below serve to demonstrate this construction in representative cases; they are not intended as benchmarks of quantum advantage.

\subsection{Main structural result}
\label{sec:main-structural-result}

The central structural result is that the cell star provides the finite-dimensional counterpart of the virtual neighbourhood required by differentiation. This follows from the support of the finite-dimensional operator and is expressed by the star-local update rule~(\ref{eq:star-local-update}). Although the same operator can be written in the matrix form ${\bf D}^{T}{\bf H}{\bf D}$,
the assembled matrix is not the primary computational object in the present framework. Instead, the operator is decomposed into the boundary and orientation relations encoded by ${\bf D}$ and ${\bf D}^{T}$, and the metric and material information encoded by the Hodge-type operator ${\bf H}$.

This separation is the point that makes the construction useful for quantum implementation. The combinatorial part of the operator can be generated from cell indices by the star-oracle, while the scalar constitutive and relaxation weights are handled separately by block encoding. The resulting circuit therefore mirrors the geometric decomposition of the finite-dimensional operator: star-oracle for local incidence data, block encoding for weights, and interference for local summation.

The construction is consequently not an ad hoc circuit for a particular equation. It gives a systematic way to compile second-order operators built from exterior derivatives and Hodge-type maps into quantum circuits once their finite-dimensional action has been expressed in star-local form.

\subsection{Circuit-level result}
\label{sec:circuit-level-result}

The quantum-circuit result is the reusable star-local architecture shown in figure~\ref{fig:qc-full}. One circuit execution implements one relaxation step. The same architecture consists of three conceptual blocks:

\[
\hbox{cyclic permutations ($\pi$)}
\longrightarrow
\hbox{star-oracle}
\longrightarrow
\hbox{block encoding}\longrightarrow
\hbox{interference-based summation}.
\]

The star-oracle generates the neighbouring cell indices and signs required by the local update. The block-encoding stage embeds the relaxation and Hodge-type scalar weights into a unitary operation. The interference stage combines the weighted local contributions into the output subspace containing the next iterate. 
The corresponding readout procedure is detailed in appendix~\ref{app:measurement_readout}.

The div--grad and curl--curl constructions differ in the degree of the cochain carrying the degrees of freedom and therefore in the corresponding star-oracle. However, the underlying circuit logic is the same. This supports the claim that the architecture is reusable across different second-order operators once their finite-dimensional action has been expressed in star-local form.

\subsection{Representative demonstrations}
\label{sec:representative-demonstrations}

We demonstrate the construction on two representative cases: a div--grad problem on a regular triangular complex and a curl--curl problem on a regular hexahedral complex. The setup of the finite-dimensional updates is given in section~\ref{sec:finite-dimensional-examples}, while the corresponding circuit construction is described in section~\ref{sec:star-local-quantum-circuit}. Additional implementation details are given in the appendices.

\begin{figure}[htbp]
    \centering
    \begin{minipage}[c]{0.32\linewidth}
        \centering
        \includegraphics[width=\linewidth]{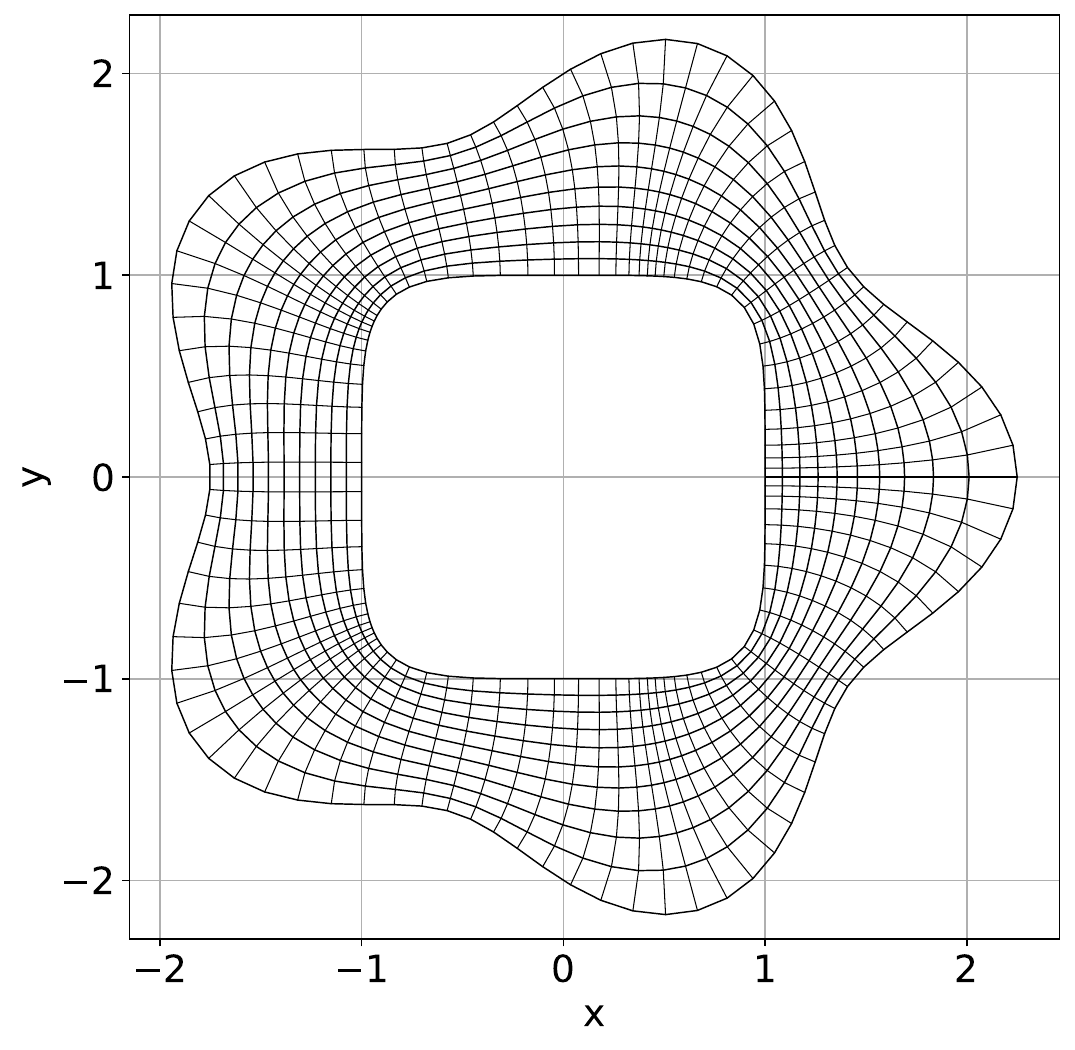}
        
        {\small (a)}
    \end{minipage}
    \hspace{0.02\linewidth}
    \begin{minipage}[c]{0.39\linewidth}
        \centering
        \includegraphics[width=\linewidth]{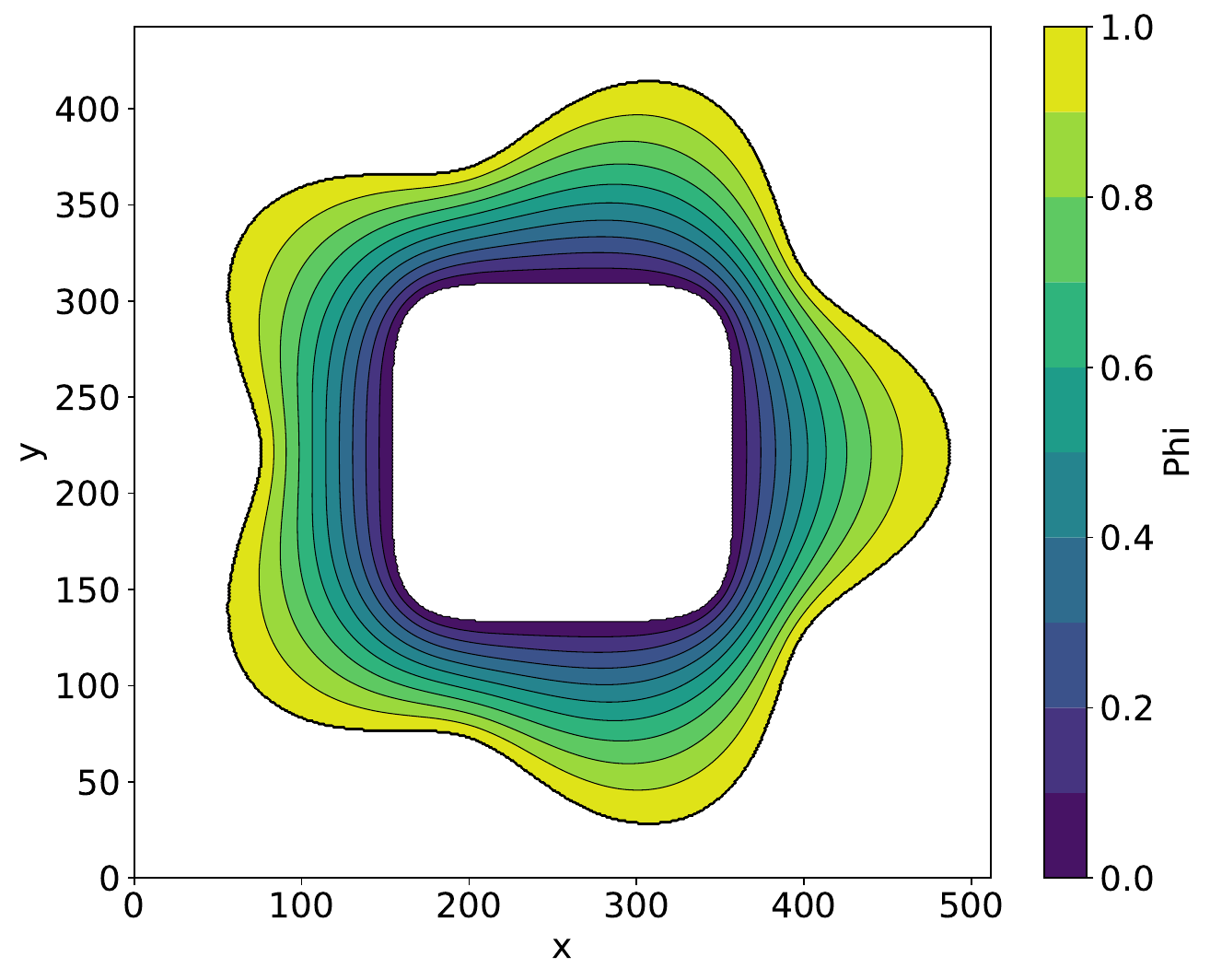}
        
        {\small (b)}
    \end{minipage}
    \caption{Div--grad demonstration. (a) Reference solution of the Laplace problem showing equipotential lines and field lines; see \cite{Geometric_solution_strategy}. (b) Solution obtained with the present star-local quantum iterative construction.}
    \label{fig:solution-of-laplace-equation}
\end{figure}

The div--grad demonstration uses a two-dimensional Laplace boundary value problem on a domain with prescribed inner and outer Dirichlet boundaries. The node-star update is implemented by the circuit shown in figure~\ref{fig:qc-divgrad}. Figure~\ref{fig:solution-of-laplace-equation} compares the resulting solution with a reference field obtained from the geometric solution strategy of \cite{Geometric_solution_strategy}. The comparison is qualitative: the purpose is to verify that the star-local circuit construction reproduces the expected equipotential and field-line structure.

\begin{figure}[htbp]
    \centering
    \begin{minipage}[c]{0.42\linewidth}
        \centering
        \includegraphics[width=\linewidth]{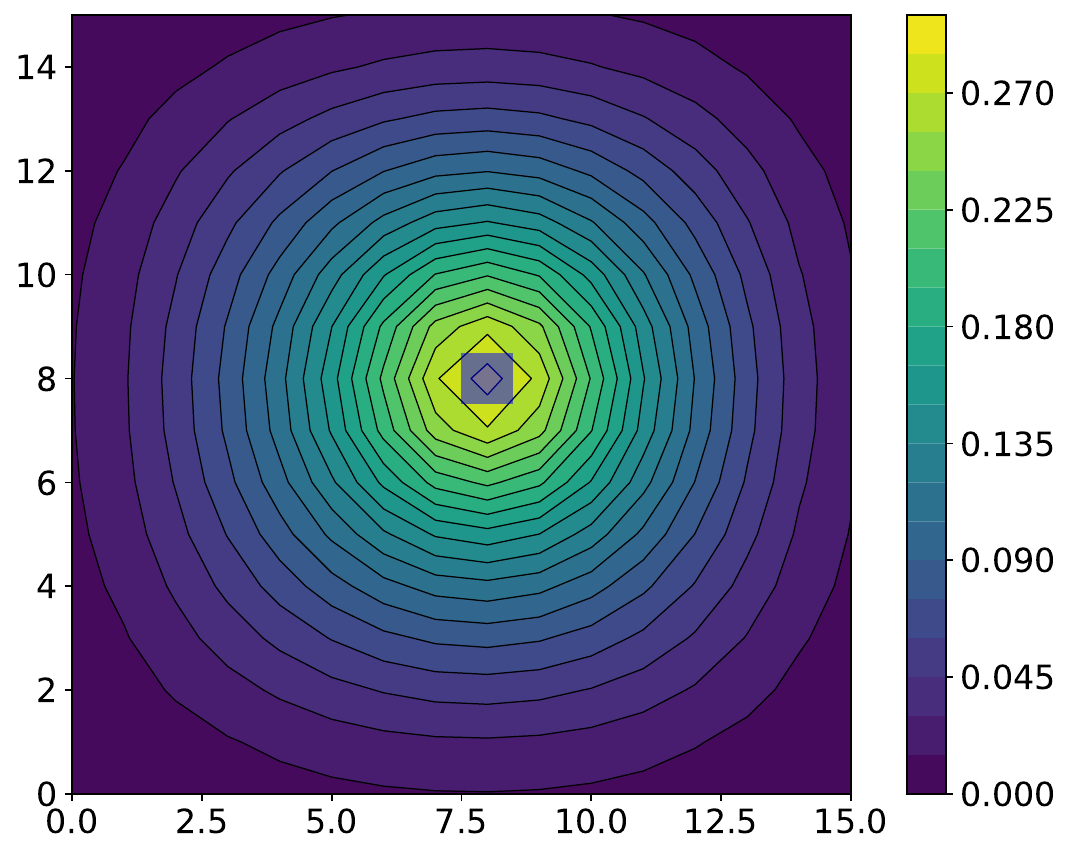}
        
        {\small (a)}
    \end{minipage}
    \hspace{0.02\linewidth}
    \begin{minipage}[c]{0.33\linewidth}
        \centering
        \includegraphics[width=\linewidth]{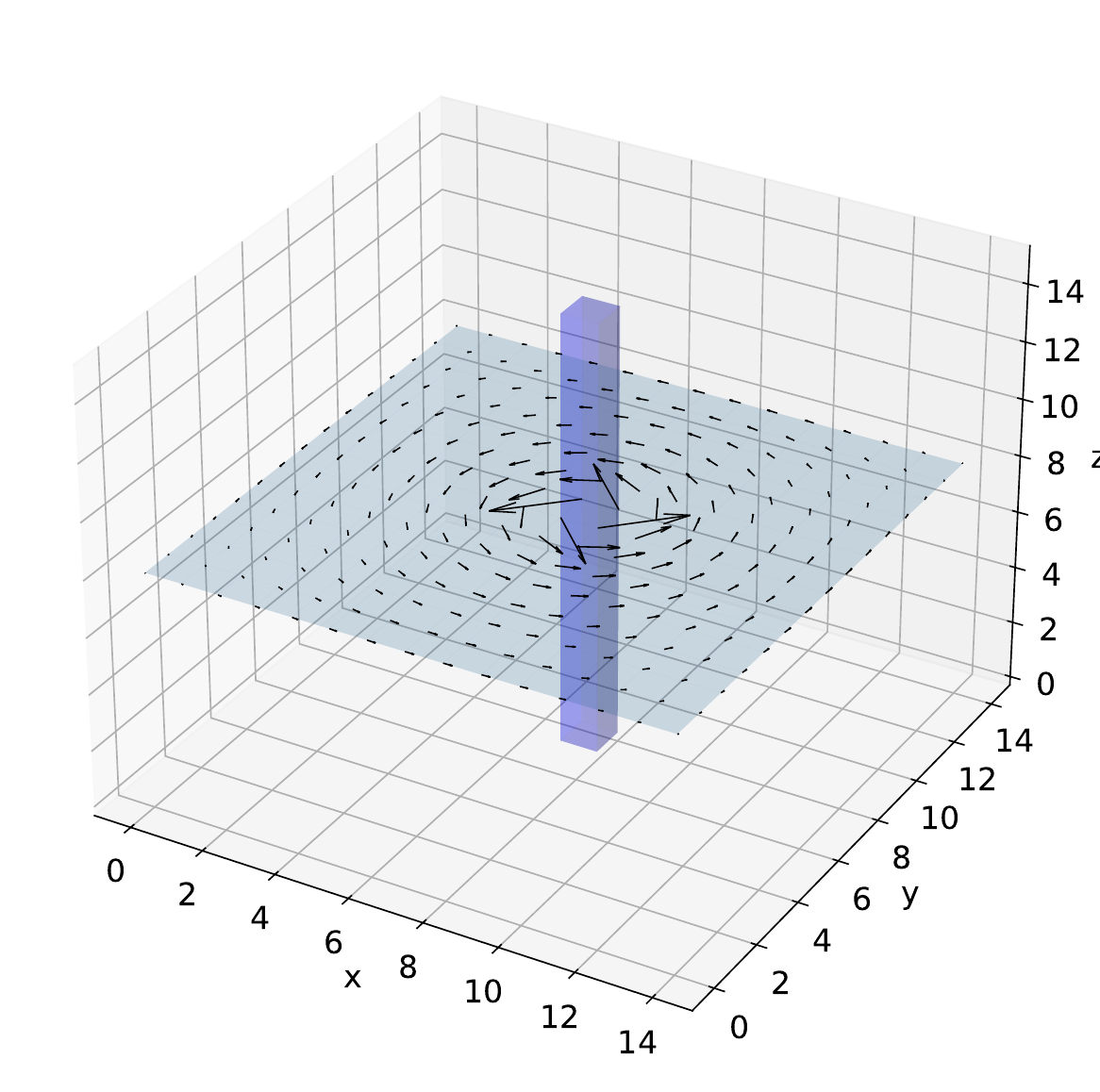}
        
        {\small (b)}
    \end{minipage}
    \caption{Curl--curl demonstration solved with the star-local quantum iterative construction. (a) Magnitude of a vector proxy of the edge cochain $a$ on the mid-plane of the cube. (b) Vector-field proxy of the derivative ${\mathrm d}a$ on the mid-plane of the cube.}
    \label{fig:example-of-curl-curl}
\end{figure}

The curl--curl demonstration uses a three-dimensional hexahedral complex with an edge-based unknown and a localized source term. The edge-star update is implemented using the curl--curl circuit architecture shown in figure~\ref{fig:qc-full}. Figure~\ref{fig:example-of-curl-curl} shows the resulting field structure on the mid-plane of the cube. The computed pattern is consistent with the expected response to a localized source: the potential-like quantity is concentrated around the source region, while the derived field structure reflects the curl--curl operator.

Together, these demonstrations show that the same construction applies when the degrees of freedom are carried by different cell types: nodes in the div--grad case and oriented edges in the curl--curl case. This is the relevant point of the examples. They demonstrate the use of the framework across different cochain degrees and local star structures, rather than serving as performance benchmarks.

\subsection{Relation to existing quantum PDE approaches}
\label{sec:relation-existing-approaches}

Most quantum algorithms for partial differential equations begin from a matrix representation of a discretized problem, for example through quantum linear-system algorithms, Hamiltonian simulation, or block encodings of sparse operators. The present approach is complementary. It does not discard matrix notation, but it does not take the assembled matrix as the fundamental object either. Instead, it starts from the local geometric structure of the differential operator and identifies the finite-dimensional neighbourhoods on which the operator acts.

This distinction is important because it changes the role of the oracle. Rather than assuming a generic sparse-matrix oracle, the circuit uses a geometrically defined star-oracle. For regularly indexed complexes, this oracle can be implemented by shifts, cyclic permutations, and phase operations. Thus the sparsity pattern is not merely stored or queried; it is generated from the cell-complex structure.

The construction also differs from quantum lattice Boltzmann approaches, although related local relaxation ideas appear there as well. In the present work, the local update is derived from the exterior-calculus representation of the differential operator and then compiled into a quantum circuit. This provides a geometric interpretation of why the update is local and how it generalizes to other cell degrees and operators.

\subsection{Scope and limitations}
\label{sec:scope-limitations}

The present implementation relies on regularly indexed complexes. This regularity is used to obtain simple closed-form star-oracles. On an unstructured complex, the same star-local finite-dimensional operator exists, but the oracle would require stored incidence data or a more involved indexing rule.

This restriction should not be interpreted as a restriction to regular Cartesian grids as physical geometries. As discussed in the Introduction and in section~\ref{sec:finite-dimensional-operator-construction}, the combinatorial structure used by the circuit and the metric structure used by the Hodge-type constitutive map play different roles. The regular indexing supplies a simple address space for the quantum circuit, while distances, material parameters, and geometric deformation enter through the Hodge-type operator.

The method also uses an encode--evolve--measure--reinitialise loop. One circuit execution implements one relaxation update, after which boundary conditions, gauge constraints, dummy-index conditions, and stopping criteria are imposed classically. This keeps the circuit architecture modular, but it also means that state preparation and readout remain important cost components. In the present work, these issues are not optimized; the focus is on the finite-dimensional operator construction and its star-local circuit implementation.

The examples are qualitative demonstrations rather than claims of quantum advantage. A full resource analysis would require optimized state preparation, hardware-aware readout, noise analysis, convergence estimates, and comparison with classical and quantum alternatives. These are important next steps.

\subsection{Conclusion and outlook}
\label{sec:conclusion-outlook}

The central conclusion is that quantum algorithms for partial differential equations can be derived directly from the mathematical structures on which the equations are built. For the second-order boundary value problems considered here, these structures are the exterior derivative, its adjoint, and Hodge-type constitutive maps. By passing from these continuous objects to finite-dimensional star-local updates, the method provides a systematic route from boundary value problems to quantum circuits. In this sense, the geometry of the differential operator is not merely a background interpretation; it becomes the organizing principle for circuit construction.

The div--grad and curl--curl demonstrations show two instances of this route. In both cases, the local circuit structure follows from the same finite-dimensional operator construction: incidence relations determine the star-oracle, Hodge-type factors determine the scalar weights, and interference performs the local summation. This supports the broader view that quantum algorithms for PDEs can be organized around the structural layers of the underlying mathematical problem rather than around a globally assembled matrix alone.

More broadly, the same viewpoint suggests extensions beyond ordinary differential forms. In particular, vector-valued and covector-valued differential forms, or more generally differential forms with values in vector bundles, would lead naturally to operators involving the covariant exterior derivative. This points toward quantum algorithms for broader classes of geometrically structured problems, including continuum mechanics and geometry processing.

Future work should address three main directions: extending the class of operators and cell complexes for which explicit star-oracles can be constructed, reducing the overhead of state preparation and readout, and developing hardware-level implementations of the encode--evolve--measure--reinitialise loop. The present work provides the finite-dimensional and circuit-level foundation for these developments.

% \funding{T.S. discloses support for the research of this work from the Finnish Ministry of Education and Culture through the Quantum Doctoral Education Pilot Program (QDOC VN/3137/2024-OKM-4) and from the Research Council of Finland through the Finnish Quantum Flagship project (JYU 359240, University of Jyväskylä). L.K. declares no relevant funding.}
% % This section is a list of funder names and grant numbers

% \roles{L.K. conceived the study and developed the mathematical and theoretical framework. T.S. developed the quantum-circuit constructions and carried out the simulations. L.K. and T.S. interpreted the results, wrote the manuscript and approved the final version.}
% % List author names and the contributions made to the article, using terms from the NISO Contributor Roles Taxonomy (CRediT) https://credit.niso.org

% \data{No datasets were generated or analysed in this study beyond the representative numerical examples.}
% For more information on IOP Publishing's research data policy see: https://publishingsupport.iopscience.iop.org/questions/research-data/

% \suppdata{Sample text inserted for demonstration.}

\appendix

% Appendix A

\section{Indexing conventions and explicit star-oracle rules}
\label{app:indexing-symmetry}

This appendix specifies the indexing conventions used by the curl--curl star-oracle. The goal is to make the map
\[
\mathcal{S}: e \mapsto \mathrm{Star}_1(e)
\]
computable from the index of the oriented edge \(e\), without accessing a stored incidence matrix.

We consider a regular hexahedral cell complex whose edges are parallel to a right-handed Cartesian coordinate system with axes \((x,y,z)\). Inner orientations of cells are encoded intrinsically by ordering their vertices; reversing the order reverses the orientation. The indexing is chosen to be equivariant under cyclic permutation of the coordinate directions. This makes it sufficient to define the oracle rule for \(x\)-directed edges; the rules for \(y\)- and \(z\)-directed edges are obtained by cyclic relabelling.

The quantum index register addresses \(2^m\) basis states. We therefore embed the physical set of indexed edges into a padded index set of size \(2^m\). Indices that do not correspond to physical degrees of freedom are treated as dummy variables and are assigned zero cochain values.

\paragraph{Cyclic symmetry.}
Let \(\pi\) denote the cyclic permutation of order three, acting on axes and on multi-indices by
\begin{equation}
\pi:(x,y,z)\mapsto(y,z,x),
\qquad
\pi:(i,j,k)\mapsto(j,k,i).
\end{equation}
By relabelling coordinate directions accordingly, \(\pi\) induces an action on each family of primal \(p\)-cells: \(x\)-edges are sent to \(y\)-edges, \(y\)-edges to \(z\)-edges, and \(z\)-edges to \(x\)-edges; likewise \(x\)-faces are sent to \(y\)-faces and then to \(z\)-faces. In this way, \(\pi\) defines an automorphism of the regular primal cell complex.

\paragraph{Equivariant labelling.}
Let \(C_p\) denote the set of primal \(p\)-cells. We use label sets
\[
L_0=\mathbb{Z}^3,\qquad
L_1=\mathbb{Z}^3\times\{x,y,z\},\qquad
L_2=\mathbb{Z}^3\times\{x,y,z\},\qquad
L_3=\mathbb{Z}^3,
\]
where the symbols \(x,y,z\) indicate the cell family. For edges, they denote the coordinate direction of the edge. For facets, they denote the coordinate direction normal to the facet. The cyclic permutation \(\pi\) acts on these label sets by
\[
\begin{aligned}
\pi\cdot(i,j,k) &= (j,k,i), 
&
\pi\cdot(i,j,k; x) &= (j,k,i; y),
\\
\pi\cdot(i,j,k; y) &= (j,k,i; z), 
&
\pi\cdot(i,j,k; z) &= (j,k,i; x).
\end{aligned}
\]
A labelling map \(\ell_p:C_p\to L_p\) is called cyclic-equivariant if
\begin{equation}
\ell_p(\pi\cdot\omega)=\pi\cdot\ell_p(\omega)
\qquad
\forall\,\omega\in C_p.
\end{equation}
Since \(\pi^3=\mathrm{Id}\), one has \((\pi\cdot)^3=\mathrm{Id}\) on both \(C_p\) and \(L_p\).

\medskip
We now specify such a cyclic-equivariant labelling of nodes, edges, facets, and volumes.

\paragraph{0-cells, or nodes.}
Let \(n_x=n_y=n_z=n\), with \(n\in\mathbb{N}\) and \(n>0\). The node set is
\begin{equation}
C_0
=
\bigl\{
\omega_0^{(i,j,k)}
\;\big|\;
0\le i\le n_x-1,\;
0\le j\le n_y-1,\;
0\le k\le n_z-1
\bigr\}.
\end{equation}
The node \(\omega_0^{(i,j,k)}\) is located at
\[
(x_0+ih,\;y_0+jh,\;z_0+kh),
\]
where \(h_x=h_y=h_z=h>0\) is the uniform spacing.

\paragraph{1-cells, or edges.}
Edges are oriented intrinsically by an ordered pair consisting of a start node and an end node. We define
\begin{equation}
\begin{aligned}
C_{1x}
&=
\bigl\{
\omega_{1x}^{(i,j,k)}
\;\big|\;
0\le i\le n_x-2,\;
0\le j\le n_y-1,\;
0\le k\le n_z-1
\bigr\},
\\
C_{1y}
&=
\bigl\{
\omega_{1y}^{(i,j,k)}
\;\big|\;
0\le i\le n_x-1,\;
0\le j\le n_y-2,\;
0\le k\le n_z-1
\bigr\},
\\
C_{1z}
&=
\bigl\{
\omega_{1z}^{(i,j,k)}
\;\big|\;
0\le i\le n_x-1,\;
0\le j\le n_y-1,\;
0\le k\le n_z-2
\bigr\}.
\end{aligned}
\end{equation}
For \((i,j,k)\) in the corresponding ranges,
\begin{equation}
\begin{aligned}
\omega_{1x}^{(i,j,k)}
&=
\bigl(
\omega_0^{(i,j,k)},
\omega_0^{(i+1,j,k)}
\bigr),
\\
\omega_{1y}^{(i,j,k)}
&=
\bigl(
\omega_0^{(i,j,k)},
\omega_0^{(i,j+1,k)}
\bigr),
\\
\omega_{1z}^{(i,j,k)}
&=
\bigl(
\omega_0^{(i,j,k)},
\omega_0^{(i,j,k+1)}
\bigr).
\end{aligned}
\end{equation}
We set
\[
C_1=C_{1x}\cup C_{1y}\cup C_{1z}.
\]

\paragraph{2-cells, or facets.}
Facets are oriented intrinsically by an ordered 4-tuple of nodes. We define
\begin{equation}
\begin{aligned}
C_{2x}
&=
\bigl\{
\omega_{2x}^{(i,j,k)}
\;\big|\;
0\le i\le n_x-1,\;
0\le j\le n_y-2,\;
0\le k\le n_z-2
\bigr\},
\\
C_{2y}
&=
\bigl\{
\omega_{2y}^{(i,j,k)}
\;\big|\;
0\le i\le n_x-2,\;
0\le j\le n_y-1,\;
0\le k\le n_z-2
\bigr\},
\\
C_{2z}
&=
\bigl\{
\omega_{2z}^{(i,j,k)}
\;\big|\;
0\le i\le n_x-2,\;
0\le j\le n_y-2,\;
0\le k\le n_z-1
\bigr\}.
\end{aligned}
\end{equation}
For \((i,j,k)\) in the corresponding ranges,
\begin{equation}
\begin{aligned}
\omega_{2x}^{(i,j,k)}
&=
\bigl(
\omega_0^{(i,j,k)},
\omega_0^{(i,j+1,k)},
\omega_0^{(i,j+1,k+1)},
\omega_0^{(i,j,k+1)}
\bigr),
\\
\omega_{2y}^{(i,j,k)}
&=
\bigl(
\omega_0^{(i,j,k)},
\omega_0^{(i,j,k+1)},
\omega_0^{(i+1,j,k+1)},
\omega_0^{(i+1,j,k)}
\bigr),
\\
\omega_{2z}^{(i,j,k)}
&=
\bigl(
\omega_0^{(i,j,k)},
\omega_0^{(i+1,j,k)},
\omega_0^{(i+1,j+1,k)},
\omega_0^{(i,j+1,k)}
\bigr).
\end{aligned}
\end{equation}
We set
\[
C_2=C_{2x}\cup C_{2y}\cup C_{2z}.
\]

\paragraph{3-cells, or volumes.}
Volumes are indexed by
\begin{equation}
C_3
=
\bigl\{
\omega_3^{(i,j,k)}
\;\big|\;
0\le i\le n_x-2,\;
0\le j\le n_y-2,\;
0\le k\le n_z-2
\bigr\}.
\end{equation}
For \((i,j,k)\) in this range, the \(3\)-cell \(\omega_3^{(i,j,k)}\) is the hexahedron with vertices
\[
\omega_0^{(i+\alpha,j+\beta,k+\gamma)},
\qquad
\alpha,\beta,\gamma\in\{0,1\},
\]
endowed with the orientation induced by the chosen vertex ordering.

\begin{figure}[htbp]
    \centering
    \includegraphics[width=0.45\textwidth]{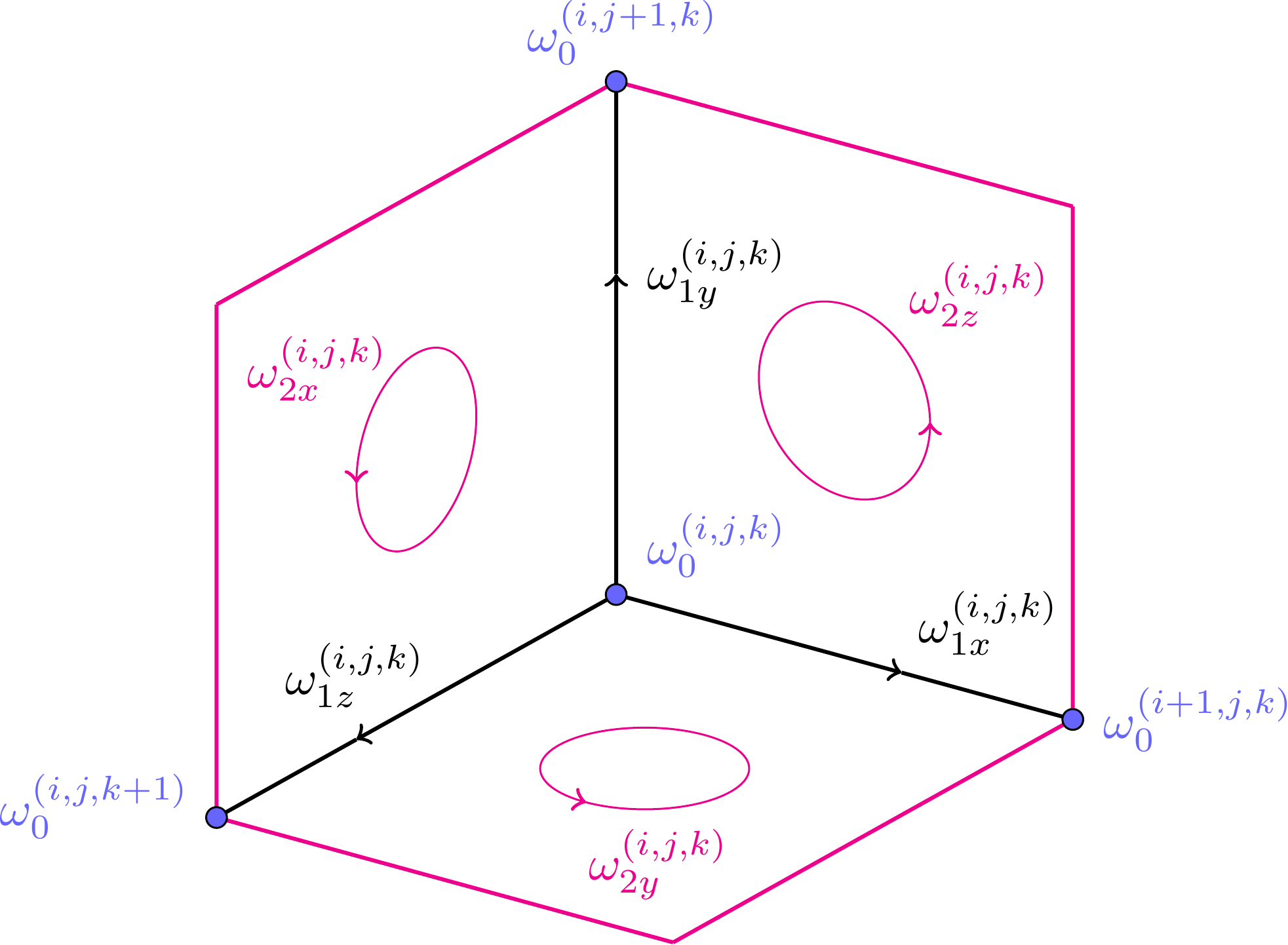}
    \caption{Indexing and orientation of \(0\)-, \(1\)-, and \(2\)-cells in a complex of regular hexahedra whose edges are parallel to a right-handed Cartesian coordinate system. The \(3\)-cell \(\omega_3^{(i,j,k)}\) is the cube whose bounding facets include, up to orientation sign, \(\omega_{2x}^{(i,j,k)}\), \(\omega_{2y}^{(i,j,k)}\), and \(\omega_{2z}^{(i,j,k)}\).}
    \label{fig:indexing_of_cells}
\end{figure}

For a graphical illustration of the indexing of the cells, see figure~\ref{fig:indexing_of_cells}.

\paragraph{Chain-level definition of the star-oracle.}
For an oriented edge \(e\in C_1\), let
\[
\mathrm{Star}(e)
=
\{f\in C_2\mid e\subset\partial f\}
\]
denote the set of oriented facets incident to \(e\). For each \(f\in\mathrm{Star}(e)\), the incidence coefficient
\[
[\partial f:e]\in\{-1,+1\}
\]
records whether the boundary orientation induced on \(e\) by \(\partial f\) agrees with the chosen orientation of \(e\).

We use this coefficient to choose the compatible orientation of the facet:
\[
\alpha_f=[\partial f:e],
\qquad
\hat f(e)=\alpha_f f .
\]
By construction, the coefficient of \(e\) in \(\partial\hat f(e)\) is \(+1\) for every \(f\in\mathrm{Star}(e)\).

Define the 2-chain-valued map
\[
F^\star(e)
=
\sum_{f\in\mathrm{Star}(e)}\hat f(e)
=
\sum_{f\in\mathrm{Star}(e)}\alpha_f f .
\]
The bounding-edge star of \(e\) is then the 1-chain
\begin{equation}
\label{s_eq:star1-def}
\mathrm{Star}_1(e)
=
\partial F^\star(e)
-
|\mathrm{Star}(e)|\,e
\in C_1 .
\end{equation}
Here \(|\mathrm{Star}(e)|\) denotes the number of oriented facets incident to \(e\). Since each oriented facet \(\hat f(e)\) has been chosen so that \(e\) appears in \(\partial\hat f(e)\) with coefficient \(+1\), this is also the number of occurrences of \(e\) in \(\partial F^\star(e)\). For an interior edge in a regular hexahedral complex, \(|\mathrm{Star}(e)|=4\).

This defines the chain-valued map
\[
\mathcal{S}:e\mapsto\mathrm{Star}_1(e),
\]
which is realised in the quantum implementation as a callable star-oracle.

\paragraph{Facet boundaries.}
The boundary operator yields
\begin{equation}
\label{eq:facet-boundaries}
\begin{aligned}
\partial\omega_{2y}^{(i,j,k)}
&=
\omega_{1z}^{(i,j,k)}
+
\omega_{1x}^{(i,j,k+1)}
-
\omega_{1z}^{(i+1,j,k)}
-
\omega_{1x}^{(i,j,k)},
\\
\partial\omega_{2z}^{(i,j,k)}
&=
\omega_{1x}^{(i,j,k)}
+
\omega_{1y}^{(i+1,j,k)}
-
\omega_{1x}^{(i,j+1,k)}
-
\omega_{1y}^{(i,j,k)}.
\end{aligned}
\end{equation}
Analogous formulas for shifted facets, such as \(\omega_{2y}^{(i,j,k-1)}\) and \(\omega_{2z}^{(i,j-1,k)}\), follow by substituting \(k\mapsto k-1\) or \(j\mapsto j-1\) in (\ref{eq:facet-boundaries}).

\paragraph{Boundary-aware formula for \(\mathrm{Star}_1(e)\).}
Let
\[
e=\omega_{1x}^{(i,j,k)} .
\]
The four candidate incident facets are
\[
\omega_{2y}^{(i,j,k)},
\qquad
\omega_{2y}^{(i,j,k-1)},
\qquad
\omega_{2z}^{(i,j,k)},
\qquad
\omega_{2z}^{(i,j-1,k)},
\]
subject to admissibility of indices. Inserting their boundaries into (\ref{s_eq:star1-def}) yields the explicit boundary-aware \(1\)-chain
\begin{equation}
\label{eq:star1-xedge}
\begin{aligned}
\mathrm{Star}_1\!\bigl(\omega_{1x}^{(i,j,k)}\bigr)
=\, &
\mathbf{1}(k\le n_z-2)
\Bigl(
-\omega_{1z}^{(i,j,k)}
-\omega_{1x}^{(i,j,k+1)}
+\omega_{1z}^{(i+1,j,k)}
\Bigr)
\\
&+
\mathbf{1}(k\ge 1)
\Bigl(
+\omega_{1z}^{(i,j,k-1)}
-\omega_{1z}^{(i+1,j,k-1)}
-\omega_{1x}^{(i,j,k-1)}
\Bigr)
\\
&+
\mathbf{1}(j\le n_y-2)
\Bigl(
+\omega_{1y}^{(i+1,j,k)}
-\omega_{1x}^{(i,j+1,k)}
-\omega_{1y}^{(i,j,k)}
\Bigr)
\\
&+
\mathbf{1}(j\ge 1)
\Bigl(
-\omega_{1x}^{(i,j-1,k)}
-\omega_{1y}^{(i+1,j-1,k)}
+\omega_{1y}^{(i,j-1,k)}
\Bigr),
\end{aligned}
\end{equation}
where \(\mathbf{1}(\cdot)\in\{0,1\}\) is a predicate equal to \(1\) when the condition holds and \(0\) otherwise. For an interior edge all four predicates equal \(1\), and (\ref{eq:star1-xedge}) reduces to a signed sum of \(4\times 3=12\) oriented neighbouring edges. This is the boundary-aware version of the formula shown in the main text.

The corresponding formulas for \(y\)- and \(z\)-directed edges are obtained by cyclic permutation:
\begin{equation}
\label{eq:star1-cyclic}
\mathrm{Star}_1\!\bigl(\omega_{1y}^{(i,j,k)}\bigr)
=
\pi\cdot
\mathrm{Star}_1\!\bigl(\omega_{1x}^{(k,i,j)}\bigr),
\qquad
\mathrm{Star}_1\!\bigl(\omega_{1z}^{(i,j,k)}\bigr)
=
\pi^2\cdot
\mathrm{Star}_1\!\bigl(\omega_{1x}^{(j,k,i)}\bigr).
\end{equation}

Equations~(\ref{eq:star1-xedge}) and (\ref{eq:star1-cyclic}) specify the classical function implemented coherently by the star-oracle in the quantum circuit. The predicates determine whether a candidate neighbouring edge is physical or should be mapped to a dummy component.

% Appendix B

\section{Measurement, readout, and offsets}
\label{app:measurement_readout}

This appendix describes how the output of one quantum relaxation step is interpreted and how the next iteration is initialized. One circuit execution produces the updated iterate in a designated output subspace. In state-vector simulation this output can be read directly from the amplitudes. On hardware, however, computational-basis measurements return probabilities and therefore do not directly reveal signed amplitudes. We describe the readout convention used in the simulations and the offset strategy used to avoid explicit sign recovery in the regularly indexed examples considered here.

\subsection{Output subspace}
\label{app:output-subspace}

After the star-oracle, block-encoding, and interference stages, the state has the schematic form
\[
\ket{\Psi_{\mathrm{out}}}
=
\ket{\mathrm{out}}
\frac{\Lambda^{s+1}}{C_s}
+
\ket{Q},
\]
where \(\ket{\mathrm{out}}\) labels the designated output subspace, \(C_s\) is a known normalization factor, and \(\ket{Q}\) collects components orthogonal to the output subspace. The vector \(\Lambda^{s+1}\) contains the cochain values corresponding to the next relaxation iterate.

In state-vector simulation, the amplitudes in the output subspace are directly available. The next iterate is reconstructed as
\[
\Lambda^{s+1}
=
C_s\,\Pi_{\mathrm{out}}\ket{\Psi_{\mathrm{out}}},
\]
where \(\Pi_{\mathrm{out}}\) denotes projection onto the designated output subspace, followed by the identification of the remaining basis amplitudes with the entries of the cochain vector.

The factor \(C_s\) depends on the normalization of the input state and on the known prefactors introduced by selector branching, block encoding, and Hadamard interference. In the simulations, these factors are tracked explicitly and applied classically after extracting the output amplitudes.

\subsection{Measurement and sign information}
\label{app:measurement-sign-information}

A computational-basis measurement of a quantum state returns samples from the probability distribution
\[
p_z = |\alpha_z|^2,
\]
where \(\alpha_z\) is the amplitude of the basis state \(\ket{z}\). Hence direct measurement does not reveal the sign of a real amplitude, nor the phase of a complex amplitude. Recovering the full set of amplitudes would in general require tomography or another amplitude-estimation strategy.

The relaxation scheme considered here requires real signed cochain values. In state-vector simulation this is not a difficulty, since the amplitudes are directly accessible. For hardware-oriented implementations, an additional strategy is required if the cochain entries can take both positive and negative values.

For the regularly indexed examples considered in this work, sign recovery can be avoided by adding offsets to the encoded cochain values. The offsets are chosen so that the values in the targeted output subspace are non-negative. The known offsets are then subtracted classically after readout.

\subsection{Offsets for \(0\)-cochains and \(1\)-cochains}
\label{app:offsets}

The offset strategy relies on the fact that adding an element in the kernel of the exterior derivative does not change the derivative. If \(\Lambda\) is a \(p\)-cochain and \(\eta\) is a \(p\)-cochain satisfying
\[
{\mathrm d}\eta = 0,
\]
then
\[
{\mathrm d}(\Lambda+\eta)
=
{\mathrm d}\Lambda .
\]
Thus \(\eta\) may be used as an offset without changing the derivative-dependent part of the field.

For the div--grad case, the unknown is a \(0\)-cochain. Adding a constant \(c\) gives
\[
{\mathrm d}(\varphi+c)
=
{\mathrm d}\varphi ,
\]
because for every oriented edge \(\omega_1\),
\[
\langle {\mathrm d}(\varphi+c),\omega_1\rangle
=
\langle \varphi+c,\partial\omega_1\rangle
=
\langle \varphi,\partial\omega_1\rangle
+
\langle c,\partial\omega_1\rangle ,
\]
and
\[
\langle c,\partial\omega_1\rangle = c-c=0 .
\]
Therefore a constant offset can be added to the scalar potential values before encoding and subtracted after readout.

For the curl--curl case, the unknown is a \(1\)-cochain. On the regularly indexed hexahedral complex with consistently oriented coordinate edges, we may add a constant directional \(1\)-cochain
\[
\eta
=
c_x\,{\mathrm d}x
+
c_y\,{\mathrm d}y
+
c_z\,{\mathrm d}z .
\]
For every oriented facet \(\omega_2\),
\[
\langle {\mathrm d}(a+\eta),\omega_2\rangle
=
\langle a+\eta,\partial\omega_2\rangle
=
\langle a,\partial\omega_2\rangle
+
\langle \eta,\partial\omega_2\rangle .
\]
The last term vanishes because the circulation of a constant directional \(1\)-cochain around the boundary of a coordinate facet is zero:
\[
\langle \eta,\partial\omega_2\rangle =0 .
\]
Hence
\[
{\mathrm d}(a+\eta)={\mathrm d}a .
\]
The constant directional offsets can therefore be used to shift the encoded edge-cochain values into a non-negative range without changing the derivative \({\mathrm d}a\).

After readout, the original cochain values are recovered by subtracting the known offsets:
\[
a_x(e)=a_x'(e)-c_x,
\qquad
a_y(e)=a_y'(e)-c_y,
\qquad
a_z(e)=a_z'(e)-c_z .
\]

\subsection{Choice of offsets}
\label{app:choice-of-offsets}

For a given iterate, offsets may be chosen so that all values in the targeted output subspace are non-negative. For example, for the curl--curl case one may choose
\[
c_x \geq -\min_e a_x(e),
\qquad
c_y \geq -\min_e a_y(e),
\qquad
c_z \geq -\min_e a_z(e),
\]
with an additional positive margin if needed. The shifted cochain is then encoded into the quantum state.

The offset changes the normalization of the encoded state. If the shifted cochain is denoted by \(a'=a+\eta\), then the prepared state is proportional to
\[
\ket{a'}
=
\frac{1}{\|a'\|}
\sum_e a'(e)\ket{e}.
\]
The normalization factor \(\|a'\|\) is tracked during state preparation. After output extraction, the known normalization and offset are removed classically.

\subsection{Encode--evolve--measure--reinitialise loop}
\label{app:encode-evolve-measure-reinitialise}

The quantum circuit implements one relaxation update. The full iterative solver is obtained by embedding this update into the outer loop
\[
\texttt{encode}
\;\longrightarrow\;
\texttt{evolve}
\;\longrightarrow\;
\texttt{measure}
\;\longrightarrow\;
\texttt{reinitialise}.
\]
In the present implementation, one iteration proceeds as follows:
\begin{enumerate}
    \item Start from a classical iterate \(\Lambda^s\) satisfying the prescribed boundary conditions, constraints, and dummy-index conditions.

    \item If offsets are used, add the corresponding constant \(0\)-cochain or constant directional \(1\)-cochain to obtain offset non-negative data for encoding.

    \item Prepare the normalized packed quantum state corresponding to the offset iterate.

    \item Execute the star-local quantum circuit for one relaxation update.

    \item Extract the amplitudes in the designated output subspace. In state-vector simulation this is done directly; in a hardware-oriented implementation this requires measurement and postprocessing.

    \item Restore the classical scale using the known normalization and circuit prefactors.

    \item If offsets were used, subtract them classically.

    \item Reimpose boundary conditions, gauge constraints, dummy-index conditions, and stopping criteria classically.

    \item Use the resulting vector as the next input iterate \(\Lambda^{s+1}\).
\end{enumerate}

The loop is repeated until a prescribed convergence criterion is satisfied, for example
\[
|\underline{\Lambda}^{s+1}-\underline{\Lambda}^{s}|
<
\mathrm{tolerance}.
\]

\subsection{Boundary conditions and constraints}
\label{app:boundary-conditions-constraints}

Boundary conditions are enforced classically between circuit executions. For Dirichlet-type data, the corresponding cochain entries are reset to their prescribed values after each relaxation step. Dummy degrees of freedom introduced by padding are set to zero.

For curl--curl problems, additional constraints may be required to select a representative of the potential-like cochain. In the present implementation, such constraints are imposed during the classical reinitialisation stage. The quantum circuit implements the local star update, while global constraints and stopping criteria are handled outside the circuit.

\subsection{Remarks on scalability of readout}
\label{app:readout-scalability}

The numerical examples in this work use state-vector simulation, which gives direct access to signed amplitudes and therefore provides a clean verification of the circuit construction. In a direct hardware implementation, the dominant practical bottlenecks are state preparation and measurement.

Full amplitude reconstruction by tomography is not scalable. The offset strategy described above is useful in the present setting because it avoids explicit sign recovery in the targeted output subspace. More general problems may require other readout strategies, such as amplitude estimation, interference with reference states, or problem-specific encodings that avoid sign ambiguities.

The present work focuses on the construction of the star-local quantum update. Optimizing state preparation, measurement, and convergence acceleration is left for future work.

\subsection{Gate Complexity}

The overall complexity of the proposed quantum algorithm is determined primarily by state preparation, shift operators, block encoding, and measurement tomography.

Exact quantum state preparation for arbitrary data generally scales exponentially with the number of qubits~\cite{deBrugiere2020}. However, approximate state-preparation methods that exploit low-entanglement structure and tensor-network representations, such as matrix product states, can substantially reduce both gate complexity and circuit depth when the data exhibit sufficient regularity~\cite{kerppo2025minimizing, Yang2025dictionarybased, Sunderhauf2024blockencoding}.

In optimized implementations, the shift operators used in the star oracle scale linearly with the number of ancillary qubits.

The block-encoding procedure scales polynomially with the number of encoded terms, and the proposed algorithm requires only a limited number of controlled \texttt{RY} gates.

In practice, the dominant computational cost typically arises from state initialization and quantum measurement (that is, state tomography), whereas the remaining algorithmic components remain asymptotically efficient. Further reductions in circuit depth and gate count can be achieved through standard quantum-circuit optimization, decomposition, and ancilla-reuse techniques~\cite{quantumoptimization, Duncan2020graphtheoretic, 10.1145/3498325, vale2023decomposition}.

% Appendix C

\section{Metric variation and local Hodge coefficients}
\label{app:metric-variation}

This appendix records how metric and material variation can be incorporated in the finite-dimensional construction. The key point is that the star-oracle used in the present circuit returns only combinatorial boundary and orientation data. Metric and material coefficients enter separately through the Hodge-type constitutive relation. Thus, if metric or material data vary locally, the oracle layer must either be supplemented by a coefficient oracle or combined with a coherent mechanism for selecting the appropriate local Hodge weights.

The underlying principle of metric variation is explained in \cite{Kettunen2014ModelingRotation}. The construction below is also a generalization of the treatment of material interfaces in generalized finite differences; see equation~(10) of \cite{BossavitPIER2001}.

\subsection{Local metric data}

Let the metric and material data be given locally on primal \(n\)-cells \(\tau_n\) through diagonal matrices
\[
\boldsymbol{\alpha}_{\tau}
=
\operatorname{diag}
\left(
\alpha_{\tau,x},
\alpha_{\tau,y},
\alpha_{\tau,z}
\right),
\]
where the diagonal entries correspond to the coordinate directions of the indexed complex. For each elementary dual contribution
\[
\tilde{\omega}_{n-p}^{\tau,j},
\]
let \(\alpha_{\tau}^{\,j}\) denote the diagonal entry of \(\boldsymbol{\alpha}_{\tau}\) selected by the direction of \(\tilde{\omega}_{n-p}^{\tau,j}\), or equivalently by the normal direction of the associated primal cell \(\omega_p^j\).

The dual \((n-p)\)-cell \(\tilde{\omega}_{n-p}\) paired with a primal \(p\)-cell \(\omega_p\) may be represented in chain form as a sum of elementary dual contributions from the neighbouring primal \(n\)-cells:
\[
\tilde{\omega}_{n-p}
=
\sum_{\tau^n \supset \omega^p}
\tilde{\omega}_{n-p}^{\tau}.
\]
Here the sum is over the primal \(n\)-cells \(\tau^n\) adjacent to the primal \(p\)-cell \(\omega^p\).

\subsection{Modified Hodge-type relation}

Applying the generalized finite-difference Hodge relation locally to each elementary dual contribution gives
\begin{equation}
\begin{split}
\langle g,\tilde{\omega}_{n-p}\rangle
&=
\sum_{\tau^n \supset \omega^p}
\langle g,\tilde{\omega}_{n-p}^{\tau}\rangle
\\
&=
\sum_{\tau^n \supset \omega^p}
\alpha_{\tau}^{\,(\omega_p)}
\frac{|\tilde{\omega}_{n-p}^{\tau}|}{|\omega_p|}
\langle f,\omega_p\rangle .
\end{split}
\label{app:eq:local-hodge-sum}
\end{equation}
Here \(\alpha_{\tau}^{\,(\omega_p)}\) denotes the diagonal entry selected by the direction associated with the primal cell \(\omega_p\).

Thus the diagonal Hodge entry associated with the primal \(p\)-cell \(\omega_p^j\) is replaced by the locally accumulated coefficient
\begin{equation}
{\bf H}_{j,j}^{\mathrm{loc}}
=
\sum_{\tau^n \supset \omega_p^j}
\alpha_{\tau}^{\,j}
\frac{|\tilde{\omega}_{n-p}^{\tau,j}|}{|\omega_p^j|}.
\label{app:eq:local-H-entry}
\end{equation}
The finite-dimensional second-order problem becomes: find the cochain values
\[
\langle \Lambda,\omega_{p-1}^i\rangle
\]
such that
\begin{equation}
\label{app:eq:finite-dhodge-variable}
\sum_j
{\bf D}^{T}_{i,j}
\left(
\sum_{\tau^n \supset \omega_p^j}
\alpha_{\tau}^{\,j}
\frac{|\tilde{\omega}_{n-p}^{\tau,j}|}{|\omega_p^j|}
\right)
\sum_k
{\bf D}_{j,k}
\langle \Lambda,\omega_{p-1}^k\rangle
=
\langle u,\tilde{\omega}_{n-p+1}^i\rangle .
\end{equation}
Equivalently, the same matrix expression
\[
{\bf D}^{T}{\bf H}{\bf D}\,\underline{\Lambda}
=
\underline{u}
\]
is retained, but the Hodge matrix entries now contain local cellwise metric and material contributions.

\subsection{Example: curl--curl operator on a hexahedral complex}

Consider the curl--curl operator on a regular hexahedral complex. Here the unknown is a \(1\)-cochain, so \(p=2\) in the constitutive relation between primal facets and dual edges. The local Hodge coefficient associated with a primal facet \(\omega_2^j\) is obtained by summing the dual-edge contributions from the primal hexahedra adjacent to that facet.

\begin{figure}[htbp]
    \centering
    \includegraphics[width=0.30\textwidth]{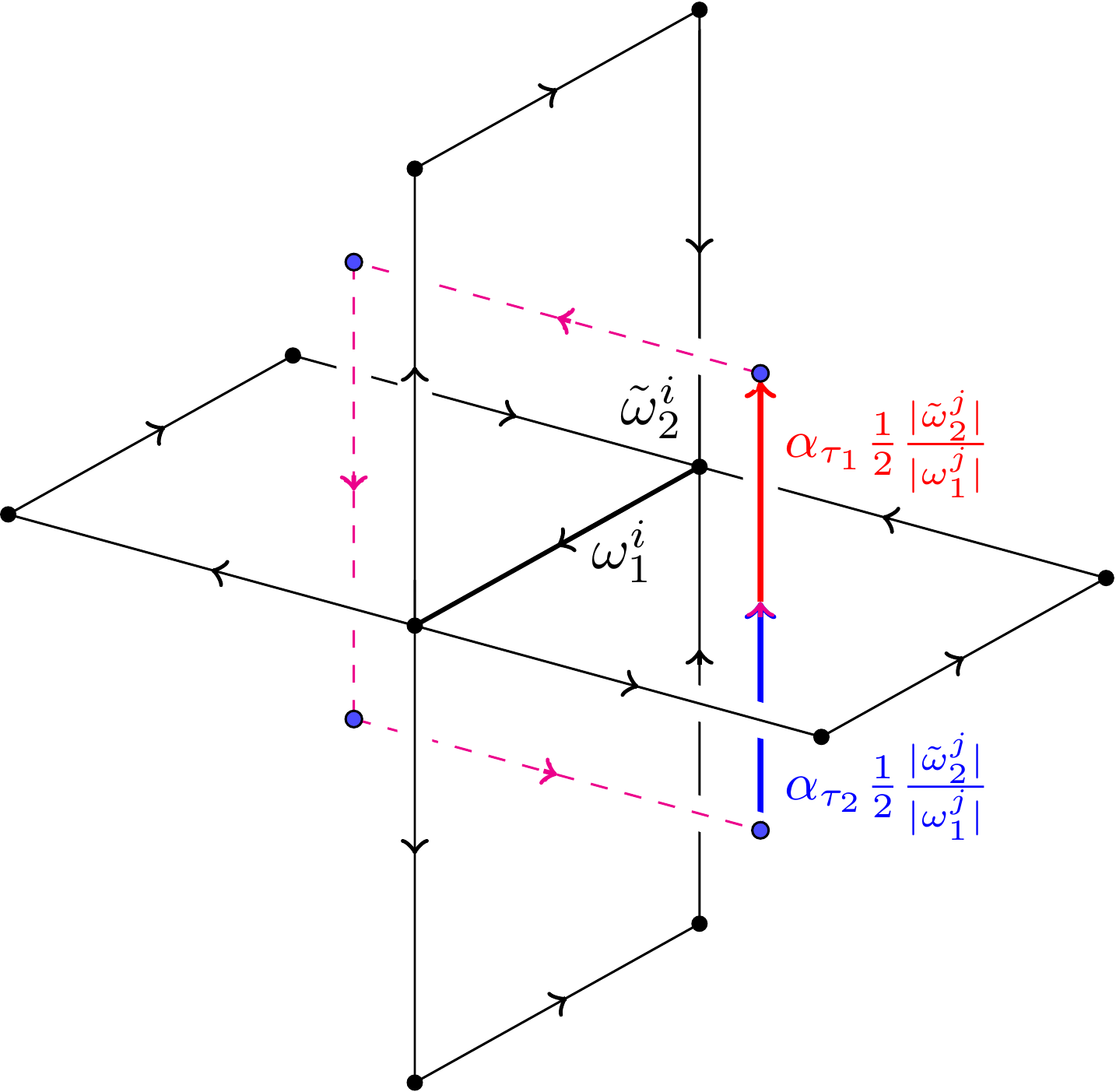}
    \caption{Decomposition of the dual contribution associated with a primal facet shared by two hexahedra. The metric or material coefficient entering the local Hodge relation is obtained from the adjacent primal \(3\)-cells.}
    \label{fig:setting_of_metric}
\end{figure}

For an interior facet shared by two hexahedra \(\tau_1\) and \(\tau_2\), equation~(\ref{app:eq:local-H-entry}) gives
\begin{equation}
\sum_{\tau^3 \supset \omega_2^j}
\alpha_{\tau}^{\,j}
\frac{|\tilde{\omega}_{1}^{\tau,j}|}{|\omega_2^j|}
=
\alpha_{\tau_1}^{\,j}
\frac{|\tilde{\omega}_{1}^{\tau_1,j}|}{|\omega_2^j|}
+
\alpha_{\tau_2}^{\,j}
\frac{|\tilde{\omega}_{1}^{\tau_2,j}|}{|\omega_2^j|}.
\label{app:eq:two-cell-hodge}
\end{equation}
For a symmetric dual construction, the elementary dual contributions each have half of the full dual-edge measure, so this becomes
\begin{equation}
\sum_{\tau^3 \supset \omega_2^j}
\alpha_{\tau}^{\,j}
\frac{|\tilde{\omega}_{1}^{\tau,j}|}{|\omega_2^j|}
=
\alpha_{\tau_1}^{\,j}
\frac{1}{2}
\frac{|\tilde{\omega}_{1}^{j}|}{|\omega_2^j|}
+
\alpha_{\tau_2}^{\,j}
\frac{1}{2}
\frac{|\tilde{\omega}_{1}^{j}|}{|\omega_2^j|}.
\label{app:eq:two-cell-hodge-half}
\end{equation}
Here \(\alpha_{\tau_1}^{\,j}\) and \(\alpha_{\tau_2}^{\,j}\) are the diagonal entries selected from \(\boldsymbol{\alpha}_{\tau_1}\) and \(\boldsymbol{\alpha}_{\tau_2}\) according to the direction of the dual edge associated with \(\omega_2^j\). The geometric decomposition is illustrated in figure~\ref{fig:setting_of_metric}.

\subsection{Implications for the quantum circuit}

The construction above shows that metric variation is compatible with a fixed indexed cell complex at the finite-dimensional level. The combinatorial star data and the metric data play different roles:

\[
\hbox{boundary and orientation data}
\quad\longrightarrow\quad
\hbox{star-oracle},
\]
\[
\hbox{metric and material data}
\quad\longrightarrow\quad
\hbox{local Hodge coefficients}.
\]

In the present quantum implementation, the star-oracle returns only the combinatorial bounding-chain data determined by the indexed incidence structure. The metric-dependent coefficients in equation~(\ref{app:eq:finite-dhodge-variable}) depend additionally on the local cellwise data \(\boldsymbol{\alpha}_{\tau}\). Therefore, to realize metric variation coherently in the quantum circuit, the oracle layer must be extended so that, together with the combinatorial star data, it also provides or enables coherent selection of the corresponding metric-dependent coefficients.

One possible implementation route is to introduce a separate coefficient oracle
\[
\mathcal{H}:
(j,\tau)
\longmapsto
\alpha_{\tau}^{\,j}
\frac{|\tilde{\omega}_{n-p}^{\tau,j}|}{|\omega_p^j|},
\]
which is queried after the star-oracle has generated the relevant local cells. The block-encoding stage would then apply the selected local Hodge weights together with the relaxation coefficients. In this way, the circuit architecture remains conceptually unchanged:
\[
\hbox{star-oracle}
\quad\longrightarrow\quad
\hbox{coefficient oracle}
\quad\longrightarrow\quad
\hbox{block encoding}
\quad\longrightarrow\quad
\hbox{interference}.
\]

The current work focuses on the case in which the star-oracle generates the local combinatorial structure and the scalar weights are simple enough to be encoded directly. Extending the implementation to fully variable metric and material data requires the additional coefficient-selection mechanism described above.

% Appendix D

\section{Numerical demonstration details}
\label{app:numerical-demonstration-details}

This appendix provides additional details for the representative div--grad and curl--curl demonstrations discussed in the main text. The purpose of these examples is illustrative and qualitative. They show how the finite-dimensional construction and the corresponding quantum circuits are instantiated for concrete boundary value problems. They are not intended as benchmarks of quantum advantage, convergence rate, or hardware performance.

\subsection{Div--grad demonstration problem}
\label{app:divgrad-demonstration-details}

The div--grad demonstration is the Laplace boundary value problem
\[
{\mathrm d}\star{\mathrm d}\phi = 0,
\]
posed on a two-dimensional domain \(\Omega\). The outer boundary is given by
\[
r = 2.0 + 0.25\cos(5\theta),
\]
and the inner boundary by
\[
|x|^5 + |y|^5 - 1 = 0.
\]
Dirichlet boundary conditions are prescribed as
\[
\phi = 1
\quad \text{on the outer boundary,}
\qquad
\phi = 0
\quad \text{on the inner boundary.}
\]

The domain is represented on a regular triangular lattice, and the corresponding relaxation step is implemented by the div--grad circuit shown in the main text. A reference solution is obtained from the geometric solution strategy cited in the main text. The comparison is used only to verify that the qualitative field structure is reproduced.

\subsection{Geometry encoding and state preparation}
\label{app:divgrad-geometry-encoding}

To initialize the div--grad demonstration, a uniform Cartesian index grid is introduced over the computational domain. Let
\[
n_x = n_y = 2^m,
\]
for some positive integer \(m\). Although the computational mesh is triangular, each grid point can be interpreted in both Cartesian and polar coordinates, allowing direct evaluation of the outer and inner boundary curves.

At each grid point \((x_i,y_j)\), one checks whether the point lies inside the outer boundary and outside the inner boundary. Points satisfying both conditions are classified as belonging to the problem domain, while the remaining points are treated as exterior or dummy degrees of freedom. In this way, the curved geometry is embedded into a structured index lattice.

The scalar field \(\phi\) is initialized from the prescribed boundary data:
\[
\phi = 1
\quad \text{on the outer boundary,}
\qquad
\phi = 0
\quad \text{on the inner excluded region.}
\]
The remaining interior values are initialized for the relaxation process.

The two-dimensional field is then flattened into the vector
\begin{equation}
\label{eq:si-phivector}
\Phi^s
=
[\phi_{0,0},\phi_{0,1},\ldots,\phi_{n_x-1,n_y-1}]^T .
\end{equation}
This vector is normalized and used for amplitude encoding in the initial state preparation of the quantum circuit.

\subsection{Explicit div--grad star rules on the triangular lattice}
\label{app:divgrad-star-rules}

For the regular triangular lattice used in the div--grad demonstration, the neighbouring nodes entering the local update depend on whether the row index is even or odd. Accordingly, the explicit function returning the node star differs in the two cases.

\paragraph{Even row \((j\) even).}
For a node \(\omega_0^{(i,j)}\), the neighbouring nodes are
\[
\begin{aligned}
\operatorname{Star}(\omega_0^{(i,j)})
={}&
\mathbf{1}(i\ge 1)\,\omega_0^{(i-1,j)}
+
\mathbf{1}(i\le n_x-2)\,\omega_0^{(i+1,j)}
\\
&+
\mathbf{1}(j\ge 1)\,\omega_0^{(i,j-1)}
+
\mathbf{1}(j\ge 1,i\ge 1)\,\omega_0^{(i-1,j-1)}
\\
&+
\mathbf{1}(j\le n_y-2)\,\omega_0^{(i,j+1)}
+
\mathbf{1}(j\le n_y-2,i\ge 1)\,\omega_0^{(i-1,j+1)} .
\end{aligned}
\]

\paragraph{Odd row \((j\) odd).}
For a node \(\omega_0^{(i,j)}\), the neighbouring nodes are
\[
\begin{aligned}
\operatorname{Star}(\omega_0^{(i,j)})
={}&
\mathbf{1}(i\ge 1)\,\omega_0^{(i-1,j)}
+
\mathbf{1}(i\le n_x-2)\,\omega_0^{(i+1,j)}
\\
&+
\mathbf{1}(j\ge 1)\,\omega_0^{(i,j-1)}
+
\mathbf{1}(j\ge 1,i\le n_x-2)\,\omega_0^{(i+1,j-1)}
\\
&+
\mathbf{1}(j\le n_y-2)\,\omega_0^{(i,j+1)}
+
\mathbf{1}(j\le n_y-2,i\le n_x-2)\,\omega_0^{(i+1,j+1)} .
\end{aligned}
\]
Here \(\mathbf{1}(\cdot)\) denotes the predicate that is equal to \(1\) when the condition is true and \(0\) otherwise. These predicates handle boundary and dummy-index cases.

These rules replace stored incidence information in the local div--grad update and determine the controlled shift pattern in the corresponding node-star oracle.

\subsection{Div--grad circuit instantiation}
\label{app:divgrad-circuit-instantiation}

For the Laplace problem with vanishing source term, the local relaxation rule takes the form
\begin{equation}
\label{eq:si-laplace-update}
\langle \varphi,\omega_0^1\rangle^{s+1}
=
\frac{\beta}{6}
\left(
\langle \varphi,\omega_0^2\rangle^s
+
\langle \varphi,\omega_0^3\rangle^s
+\cdots+
\langle \varphi,\omega_0^7\rangle^s
\right)
+
(1-\beta)\langle \varphi,\omega_0^1\rangle^s .
\end{equation}
This is the node-star update implemented by the div--grad circuit.

The state initialized into the quantum circuit is the normalized vector \(\Phi^s\) of equation~(\ref{eq:si-phivector}). Since its dimension is
\[
n=n_xn_y=2^{2m},
\]
the number of qubits required to encode \(\Phi^s\) is
\[
N=\log_2 n=2m .
\]
In addition to this \(N\)-qubit state register, the div--grad circuit uses a selector register controlling the computational branches and one ancilla qubit for block encoding.

Because the triangular lattice distinguishes even and odd rows, the circuit requires one additional control condition in the star-oracle stage to select the correct neighbour pattern. Operationally, this amounts to conditioning the left- and right-shift operations on the parity of the row-index register. Apart from this parity-dependent routing, the block-encoding and interference stages follow the same principles as in the curl--curl construction.

\subsection{Curl--curl demonstration problem}
\label{app:curlcurl-demonstration-details}

The curl--curl demonstration is posed on a regular three-dimensional hexahedral complex covering a cubic domain. The governing equation is
\[
{\mathrm d}\star{\mathrm d}a = j
\qquad
\text{in }\Omega,
\]
where \(a\) is a \(1\)-form and \(j\) is a prescribed source term.

A localized excitation is introduced by assigning a nonzero source term to central vertically oriented edges of the domain, while \(j=0\) elsewhere. Homogeneous Dirichlet boundary conditions are imposed on the outer boundary.

The derivative \({\mathrm d}a\) is evaluated on the regular hexahedral complex, and the numerical results are examined on the mid-plane of the cube. In the main text, the magnitude of a vector proxy of \(a\) and the vector-field proxy of \({\mathrm d}a\) are shown. The expected qualitative behaviour is the formation of circulation around the source region, characteristic of a curl--curl-type problem with a localized excitation.

\subsection{Purpose of the demonstrations}
\label{app:purpose-demonstrations}

Both demonstrations use the same algorithmic structure: a star-oracle generates local incidence data, block encoding applies scalar relaxation weights, and interference-based summation forms the next iterate. The div--grad case demonstrates the construction for node-based \(0\)-cochains, while the curl--curl case demonstrates it for edge-based \(1\)-cochains.

The examples are therefore intended to verify the implementation of the star-local construction across different cochain degrees and local star structures. They are not intended to provide a full numerical error analysis or a resource comparison with classical solvers.

\bibliographystyle{iopart-num}
\bibliography{qst_references}

@article{AlbaneseRubinacci1988,
  author  = {Albanese, R. and Rubinacci, G.},
  title   = {Integral formulation for 3D eddy-current computation using edge elements},
  journal = {IEE Proceedings A (Physical Science, Measurement and Instrumentation)},
  volume  = {135},
  number  = {7},
  pages   = {457--463},
  year    = {1988}
}

@book{Baez-Muniain,
  author    = {Baez, John C. and Muniain, Javier P.},
  title     = {Gauge Fields, Knots and Gravity},
  series    = {Series on Knots and Everything},
  volume    = {4},
  publisher = {World Scientific},
  address   = {Singapore},
  year      = {1994}
}

@misc{bastida2026quantum,
  author        = {Bastida-Zamora, A. and Budinski, L. and Kerppo, O. and Lahtinen, V. and Niemimäki, O. and Steadman, W. and Zamora-Zamora, R. and Sagaut, P. and Bohun, V. and Koch-Janusz, M. and others},
  title         = {Quantum algorithm for the lattice Boltzmann method with applications on real quantum devices},
  year          = {2026},
  archivePrefix = {arXiv},
  eprint        = {2603.02127},
  primaryClass  = {quant-ph}
}

@book{Bossavit-book,
  author     = {Alain Bossavit},
  title         = {Computational Electromagnetism: Variational Formulations, Complementarity, Edge Elements},
  publisher = {Academic Press},
  address   = {San Diego},
  year        = {1997},
  pages     = {384}
}

@article{Bossavit1998,
  author  = {Bossavit, Alain},
  title   = {Magnetostatic problems in multiply connected regions: properties of the curl operator},
  journal = {IEE Proceedings A},
  volume  = {135},
  number  = {3},
  pages   = {179--187},
  year    = {1988}
}

@article{Bossavit-Japan2,
  author  = {Bossavit, Alain},
  title   = {On the geometry of electromagnetism, (2): Geometrical objects},
  journal = {Journal of Japan Society of Applied Electromagnetism and Mechanics},
  volume  = {6},
  pages   = {114--123},
  year    = {1998}
}

@article{Bossavit-Kettunen1999,
  author  = {Bossavit, A. and Kettunen, L.},
  title   = {Yee-like schemes on a tetrahedral mesh, with diagonal lumping},
  journal = {International Journal of Numerical Modelling: Electronic Networks, Devices and Fields},
  volume  = {12},
  pages   = {129--142},
  year    = {1999}
}

@incollection{BossavitPIER2001,
  author    = {Bossavit, Alain},
  title     = {{``Generalized finite differences'' in computational electromagnetics}},
  editor    = {Teixeira, Fernando L.},
  booktitle = {Progress in Electromagnetics Research, PIER},
  volume    = {32},
  pages     = {45--64},
  publisher = {EMW},
  address   = {Cambridge, MA},
  year      = {2001}
}

@article{Budinski2021ADE,
  author  = {Budinski, Luka},
  title   = {Quantum algorithm for the advection--diffusion equation simulated with the lattice Boltzmann method},
  journal = {Quantum Information Processing},
  volume  = {20},
  number  = {2},
  pages   = {57},
  year    = {2021},
  doi     = {10.1007/s11128-021-02996-3}
}

@article{budinski2023efficient,
  author  = {Budinski, Luka and Niemimäki, Olli and Zamora-Zamora, Ricardo and Lahtinen, Ville},
  title   = {Efficient parallelization of quantum basis state shift},
  journal = {Quantum Science and Technology},
  volume  = {8},
  number  = {4},
  pages   = {045031},
  year    = {2023},
  doi     = {10.1088/2058-9565/acfab7}
}

@misc{camps2022fable,
  author        = {Camps, D. and Van Beeumen, R.},
  title         = {{FABLE}: Fast Approximate Quantum Circuits for Block-Encodings},
  year          = {2022},
  archivePrefix = {arXiv},
  eprint        = {2205.00081},
  primaryClass  = {quant-ph}
}

@article{camps2023explicit,
  author  = {Camps, Daan and Lin, Lin and Van Beeumen, Roel and Yang, Chao},
  title   = {Explicit quantum circuits for block encodings of certain sparse matrices},
  journal = {SIAM Journal on Matrix Analysis and Applications},
  volume  = {45},
  number  = {1},
  pages   = {801--827},
  year    = {2024},
  doi     = {10.1137/22M1484298},
  eprint  = {2203.10236},
  archivePrefix = {arXiv},
  primaryClass = {quant-ph}
}

@article{Childs2017QLSA,
  author  = {Childs, Andrew M. and Kothari, Robin and Somma, Rolando D.},
  title   = {Quantum algorithm for systems of linear equations with exponentially improved dependence on precision},
  journal = {SIAM Journal on Computing},
  volume  = {46},
  pages   = {1920--1950},
  year    = {2017},
  doi     = {10.1137/16M1087072}
}

@article{Childs2021PDE,
  author  = {Childs, Andrew M. and Liu, Jin-Peng and Ostrander, Aaron},
  title   = {High-precision quantum algorithms for partial differential equations},
  journal = {Quantum},
  volume  = {5},
  pages   = {574},
  year    = {2021},
  doi     = {10.22331/q-2021-11-10-574}
}

@book{Ciarlet1978,
  author    = {Ciarlet, Philippe G.},
  title     = {The Finite Element Method for Elliptic Problems},
  publisher = {North-Holland},
  address   = {Amsterdam},
  year      = {1978},
  pages     = {529}
}

@article{Clader2013,
  author  = {Clader, B. D. and Jacobs, B. C. and Sprouse, C. R.},
  title   = {Preconditioned quantum linear system algorithm},
  journal = {Physical Review Letters},
  volume  = {110},
  pages   = {250504},
  year    = {2013},
  doi     = {10.1103/PhysRevLett.110.250504}
}

@misc{Costa2022Review,
  author        = {Morales, Mauro E. S. and Pira, Lirandë and Schleich, Philipp and Koor, Kelvin and Costa, Pedro C. S. and An, Dong and Aspuru-Guzik, Alán and Lin, Lin and Rebentrost, Patrick and Berry, Dominic W.},
  title         = {Quantum linear systems algorithms: a survey of algorithms and applications},
  year          = {2023},
  archivePrefix = {arXiv},
  eprint        = {2305.01199},
  primaryClass  = {quant-ph}
}

@phdthesis{deBrugiere2020,
  author = {de Brugière, Timothée Goubault},
  title  = {Methods for optimizing the synthesis of quantum circuits},
  school = {Université Paris-Saclay},
  year   = {2020}
}

@article{Duncan2020graphtheoretic,
  author  = {Duncan, R. and Kissinger, A. and Perdrix, S. and van de Wetering, J.},
  title   = {Graph-theoretic simplification of quantum circuits with the ZX-calculus},
  journal = {Quantum},
  volume  = {4},
  pages   = {279},
  year    = {2020},
  doi     = {10.22331/q-2020-06-04-279}
}

@book{Flanders,
  author     = {Harley Flanders},
  title          = {Differential Forms with Applications to the Physical Sciences},
  publisher = {Dover Publications},
  address   = {New York},
  year         = {1989},
  edition     = {Revised edition},
  pages      = {205}
}

@book{Frankel,
  author    = {Frankel, Theodore},
  title     = {The Geometry of Physics: An Introduction},
  edition   = {3},
  publisher = {Cambridge University Press},
  address   = {Cambridge},
  year      = {2012}
}

@article{HHL2009,
  author  = {Harrow, Aram W. and Hassidim, Avinatan and Lloyd, Seth},
  title   = {Quantum algorithm for linear systems of equations},
  journal = {Physical Review Letters},
  volume  = {103},
  pages   = {150502},
  year    = {2009},
  doi     = {10.1103/PhysRevLett.103.150502}
}

@phdthesis{Hirani2003,
  author  = {Hirani, Anil N.},
  title   = {Discrete exterior calculus},
  school  = {California Institute of Technology},
  address = {Pasadena, CA},
  year    = {2003},
  month   = may
}

@article{Hodge1934,
  author  = {Hodge, W. V. D.},
  title   = {A Dirichlet problem for harmonic functionals, with applications to analytic varieties},
  journal = {Proceedings of the London Mathematical Society},
  volume  = {36},
  pages   = {257--303},
  year    = {1934}
}

@book{Hodge1941,
  author    = {Hodge, W. V. D.},
  title     = {The Theory and Applications of Harmonic Integrals},
  publisher = {Cambridge University Press},
  address   = {Cambridge},
  year      = {1941}
}

@article{PhysRevA.93.032318,
  author  = {Iten, R. and Colbeck, R. and Kukuljan, I. and Home, J. and Christandl, M.},
  title   = {Quantum circuits for isometries},
  journal = {Physical Review A},
  volume  = {93},
  pages   = {032318},
  year    = {2016},
  doi     = {10.1103/PhysRevA.93.032318}
}

@article{10.1145/3498325,
  author  = {Iten, R. and Moyard, R. and Metger, T. and Sutter, D. and Woerner, S.},
  title   = {Exact and practical pattern matching for quantum circuit optimization},
  journal = {ACM Transactions on Quantum Computing},
  volume  = {3},
  number  = {1},
  year    = {2022},
  doi     = {10.1145/3498325}
}

@article{quantumoptimization,
  author  = {Karuppasamy, K. and Puram, V. and Johnson, S. and Thomas, J. P.},
  title   = {A comprehensive review of quantum circuit optimization: Current trends and future directions},
  journal = {Quantum Reports},
  volume  = {7},
  number  = {1},
  year    = {2025},
  doi     = {10.3390/quantum7010002}
}

@misc{kerppo2025minimizing,
  author        = {Kerppo, O. and Steadman, W. and Niemimäki, O. and Lahtinen, V.},
  title         = {Minimizing entanglement entropy for enhanced quantum state preparation},
  year          = {2025},
  archivePrefix = {arXiv},
  eprint        = {2507.22562},
  primaryClass  = {quant-ph}
}

@article{Kettunen2014ModelingRotation,
  author  = {Kettunen, L. and Kurz, S. and Tarhasaari, T. and Räisänen, V. and Stenvall, A. and Suuriniemi, S.},
  title   = {Modeling rotation in electrical machines},
  journal = {IEEE Transactions on Magnetics},
  volume  = {50},
  number  = {4},
  pages   = {1--10},
  year    = {2014},
  doi     = {10.1109/TMAG.2013.2290101}
}

@book{Krueger2016,
  author    = {Krueger, T. and Kusumaatmaja, H. and Kuzmin, A. and Shardt, O. and Silva, G. and Viggen, E.},
  title     = {The Lattice Boltzmann Method: Principles and Practice},
  series    = {Graduate Texts in Physics},
  publisher = {Springer},
  address   = {Cham},
  year      = {2016}
}

@book{Mohammed2019,
  author    = {Mohamad, A. A.},
  title     = {Lattice Boltzmann Method: Fundamentals and Engineering Applications with Computer Codes},
  publisher = {Springer},
  address   = {London},
  year      = {2019}
}

@article{Montanaro2016,
  author  = {Montanaro, Ashley},
  title   = {Quantum algorithms: an overview},
  journal = {npj Quantum Information},
  volume  = {2},
  pages   = {15023},
  year    = {2016},
  doi     = {10.1038/npjqi.2015.23}
}

@book{Munkres1984,
  author    = {Munkres, James R.},
  title     = {Elements of Algebraic Topology},
  publisher = {Addison-Wesley},
  address   = {Menlo Park, CA},
  year      = {1984}
}

@article{Pellikka2010,
  author  = {Pellikka, M. and Suuriniemi, S. and Kettunen, L.},
  title   = {Homology in Electromagnetic Boundary Value Problems},
  journal = {Boundary Value Problems},
  volume  = {2010},
  pages   = {381953},
  year    = {2010},
  doi     = {10.1155/2010/381953}
}

@article{Pellikka2013,
  author  = {Pellikka, M. and Suuriniemi, S. and Kettunen, L. and Geuzaine, C.},
  title   = {Homology and Cohomology Computation in Finite Element Modeling},
  journal = {SIAM Journal on Scientific Computing},
  volume  = {35},
  number  = {5},
  pages   = {1195--1214},
  year    = {2013}
}

@article{Geometric_solution_strategy,
  author  = {Poutala, A. and Tarhasaari, T. and Kettunen, L.},
  title   = {Geometric solution strategy of Laplace problems with free boundary},
  journal = {International Journal for Numerical Methods in Engineering},
  volume  = {105},
  number  = {10},
  pages   = {723--746},
  year    = {2016},
  doi     = {10.1002/nme.4988}
}

@article{Shakeel_2020,
  author  = {Shakeel, Asif},
  title   = {Efficient and scalable quantum walk algorithms via the quantum Fourier transform},
  journal = {Quantum Information Processing},
  volume  = {19},
  number  = {9},
  year    = {2020},
  doi     = {10.1007/s11128-020-02834-y}
}

@article{Shende_quantum_state,
  author  = {Shende, Vivek V. and Bullock, Stephen S. and Markov, Igor L.},
  title   = {Synthesis of quantum-logic circuits},
  journal = {IEEE Transactions on Computer-Aided Design of Integrated Circuits and Systems},
  volume  = {25},
  number  = {6},
  pages   = {1000--1010},
  year    = {2006},
  doi     = {10.1109/TCAD.2005.855930}
}

@article{Sunderhauf2024blockencoding,
  author  = {Sünderhauf, C. and Campbell, E. and Camps, J.},
  title   = {Block-encoding structured matrices for data input in quantum computing},
  journal = {Quantum},
  volume  = {8},
  pages   = {1226},
  year    = {2024},
  doi     = {10.22331/q-2024-01-11-1226}
}

@article{Suuriniemi2002,
  author  = {Suuriniemi, S. and Tarhasaari, T. and Kettunen, L.},
  title   = {Generalization of the spanning-tree technique},
  journal = {IEEE Transactions on Magnetics},
  volume  = {38},
  number  = {2},
  pages   = {525--528},
  year    = {2002},
  doi     = {10.1109/20.996139}
}

@phdthesis{Suuriniemi2004,
  author    = {Suuriniemi, S.},
  title     = {Homological Computations in Electromagnetic Modeling},
  school    = {Tampere University of Technology, Department of Electrical Engineering},
  address   = {Tampere, Finland},
  year      = {2004},
  type      = {Ph.D. thesis}
}

@article{Tarhasaari-Kettunen-Bossavit1999,
  author  = {Tarhasaari, T. and Kettunen, L. and Bossavit, A.},
  title   = {Some realizations of a discrete Hodge operator: A reinterpretation of finite element techniques},
  journal = {IEEE Transactions on Magnetics},
  volume  = {35},
  number  = {3},
  pages   = {1494--1497},
  year    = {1999}
}

@misc{vale2023decomposition,
  author        = {Vale, R. and Azevedo, T. M. D. and Araújo, I. and Araujo, I. F. and Silva, A. J.},
  title         = {Decomposition of multi-controlled special unitary single-qubit gates},
  year          = {2023},
  archivePrefix = {arXiv},
  eprint        = {2302.06377},
  primaryClass  = {quant-ph}
}

@article{Venegas_Andraca_2012,
  author  = {Venegas-Andraca, Salvador E.},
  title   = {Quantum walks: a comprehensive review},
  journal = {Quantum Information Processing},
  volume  = {11},
  number  = {5},
  pages   = {1015--1106},
  year    = {2012},
  doi     = {10.1007/s11128-012-0432-5}
}

@article{Weiland1984,
title = "On the Numerical Solution of {M}axwell's Equations and Applications in the Field of Accelerator Physics",
journal = "Particle Accelerator",
author = "T. Weiland",
year = "1984",
volume = "15",
pages = "245--292"
}

@book{Whitney1957,
  author    = {Whitney, Hassler},
  title     = {Geometric Integration Theory},
  publisher = {Princeton University Press},
  address   = {Princeton, NJ},
  year      = {1957}
}

@article{Yang2025dictionarybased,
  author  = {Yang, C. and Li, Z. and Yao, H. and Fan, Z. and Zhang, G. and Liu, J.},
  title   = {Dictionary-based Block Encoding of Sparse Matrices with Low Subnormalization and Circuit Depth},
  journal = {Quantum},
  volume  = {9},
  pages   = {1805},
  year    = {2025},
  doi     = {10.22331/q-2025-07-22-1805}
}

@article{Yee1966,
  author  = {Yee, Kane},
  title   = {Numerical solution of initial boundary value problems involving Maxwell's equations in isotropic media},
  journal = {IEEE Transactions on Antennas and Propagation},
  volume  = {14},
  pages   = {302--307},
  year    = {1966},
  doi     = {10.1109/TAP.1966.1138693}
}

\end{document}